\documentclass{aa}
\usepackage[dvipsnames]{xcolor} 
\usepackage{graphicx, subfigure, amsmath, pifont}
\usepackage{amssymb}
\usepackage[breaklinks=true]{hyperref} 
\hypersetup{colorlinks=true,citecolor=Blue}
\usepackage{natbib}
\bibpunct{(}{)}{;}{a}{}{,} 

\newcommand{\cmark}{\ding{51}}%
\newcommand{\xmark}{\ding{55}}%
\usepackage[english]{babel}
\usepackage{blindtext}
\usepackage[normalem]{ulem}

\begin{document} 

 \title{Hints on the origins of particle traps in protoplanetary disks given by the $M_{\rm{dust}}-M_{\star}$ relation}

   \author{
   Paola Pinilla\inst{1},
   Ilaria Pascucci\inst{2,3}, and
   Sebastian Marino \inst{1}
   }

   \institute{Max-Planck-Institut f\"{u}r Astronomie, K\"{o}nigstuhl 17, 69117, Heidelberg, Germany, \email{pinilla@mpia.de}
    \and Lunar and Planetary Laboratory, The University of Arizona, Tucson, AZ 85721, USA
    \and Earths in Other Solar Systems Team, NASA Nexus for Exoplanet System Science
     }
   \date{}

 
  \abstract
   {Demographic surveys of protoplanetary disks, carried out mainly with the Atacama Large Millimeter/submillimete Array, have provided access to a large range of disk dust masses ($M_{\rm{dust}}$) around stars with different stellar types and in different star-forming regions. These surveys found a power-law relation between $M_{\rm{dust}}$ and $M_{\star}$ that steepens in time, but which is also flatter for transition disks (TDs).}  
   {We aim to study the effect of dust evolution in the  $M_{\rm{dust}}-M_{\star}$ relation. In particular, we are interested in investigating the effect of particle traps on this relation. }
   {We performed dust evolution models, which included perturbations to the gas surface density with different amplitudes to investigate the effect of particle trapping on the $M_{\rm{dust}}-M_{\star}$ relation. These perturbations were aimed at mimicking pressure bumps that originated from planets. We focused on the effect caused by different stellar and disk masses based on exoplanet statistics that demonstrate a dependence of planet mass on stellar mass and metallicity.} 
   {Models of dust evolution can reproduce the observed $M_{\rm{dust}}-M_{\star}$ relation in different star-forming regions when strong pressure bumps are included and when the disk mass scales with stellar mass (case of $M_{\rm{disk}}=0.05\,M_\star$ in our models). This result arises from dust trapping and dust growth beyond centimeter-sized grains inside pressure bumps. However, the flatter relation of $M_{\rm{dust}}-M_{\star}$ for TDs and disks with substructures cannot be reproduced by the models unless the formation of boulders is inhibited inside pressure bumps. }
  {In the context of  pressure bumps originating from planets, our results agree with current exoplanet statistics on giant planet occurrence increasing with stellar mass, but we cannot draw a conclusion about the type of planets needed in the case of low-mass stars. This is attributed to the fact that for $M_\star<1\,M_\odot$, the observed $M_{\rm{dust}}$ obtained from models is very low due to the efficient growth of dust particles beyond centimeter-sizes inside pressure bumps. }


   \titlerunning{$M_{\rm{dust}}-M_{\star}$ Relation and Particle Traps}
    \maketitle

%
\section{Introduction}                  \label{sect:intro}

Censuses performed on the thousands of exoplanets that have been discovered thus far have revealed a large diversity of planetary architectures, along with a variety of  trends that may be the result of the common physical processes ruling the formation and evolution of planets in disks around young stars. Exoplanet statistics show that giant planet occurrence increases with stellar metallicity and mass \citep[e.g.,][]{fischer2005, johnson2010}, while sub-Neptune type planets are more common around M-dwarfs than around Sun-like stars \citep[e.g.,][]{mulders2018}. These observed trends stand as open challenges for current theories of planet formation, examining  stellar and disk properties; in particular, questioning how the effects of the ratio between stellar mass and disk mass affect the final outcome of planet formation.  

Current observations of protoplanetary disks with the Atacama Large Millimeter/submillimete Array (ALMA) provide demographic surveys of  disks \citep{ansdell2016,ansdell2017, ansdell2018, barenfeld2016, barenfeld2017, pascucci2016, cazzoletti2019, cieza2019, long2019, terwisga2019} that aid in the understanding of the missing keys for current planet formation models. Using continuum observations of disks, we can currently access to different disk properties, including a large range of disk dust masses - an important property to understand the amount of material available in young disks to form different types of planets. The most common method to determine disk mass is to assume that the dust emission is optically thin over most of the disk volume and, therefore, that the detected flux is proportional to the dust disk mass \citep[e.g.,][]{beckwith1990}. This dust mass is usually used as a tracer of the total disk mass by assuming a constant dust-to-gas ratio,  conventionally the interstellar medium ratio, that is, 1/100. 

Obtaining information about the amount of gas available in disks is challenging because faint optically thin molecular lines, such as $^{13}$CO and C$^{18}$O, must be observed and their intensity can be also affected by isotope selective processes \citep{bruderer2012, miotello2014, miotello2016} and the unknown carbon abundance relative to H$_2$ \citep[e.g.,][]{kama2016, schwarz2016}. In addition, even $^{13}$CO and C$^{18}$O may be optically thick inside the CO ice line at the disk midplane, resulting in an underestimation of the disk mass \citep{zhang2017}, especially for warm disks around Herbig stars. In these cases, observations of rare CO isotopologues, such as $^{13}$C$^{17}$O are needed (although  challenging) for a more robust estimation of gas disk masses \citep{booth2019} and, therefore, the best knowledge that we have so far about the disk material still comes from the dust emission. 

Information obtained about the dusty material in disks from observations is, however, prone to uncertainties as well since it is based on several assumptions, such as the disk temperature (which is typically assumed to be 20\,K but could vary depending on disk size; \citealt{hendler2017}), and the dust opacity. The dust opacity is the major source of uncertainty because it depends on the grain size and composition, which are unknown in protoplanetary disks \citep[e.g.,][]{birnstiel2018}.  Nevertheless, when comparing millimeter-observations of protoplanetary disks with models of disk evolution, the same assumptions can be taken in order to facilitate the comparison and understand the effect of different crucial physical processes in the formation of planets, such as grain growth and drift. 

Due to the fast radial drift that millimeter- and centimeter-sized particles undergo in the outer parts ($>$20\,au) of protoplanetary disks, models of dust evolution at million-year timescales contradict millimeter observations of disks \citep{birnstiel2010, pinilla2012}. This is because the models predict a millimeter spectral index (an indicator of grain growth) that is higher than that which is observed for most  protoplanetary disks in different star forming regions \citep[e.g.,][]{ricci2010, testi2014}. Reducing or completing the suppression of the radial drift of dust particles, for instance, with the presence of particle traps or pressure bumps, is necessary in order to keep millimeter grains in protoplanetary disks for million-year timescales \citep{pinilla2012}. 

Observational surveys that measure the disk dust mass found a power-law relation between $M_{\rm{dust}}$ and $M_{\star}$ that steepens with time \citep[e.g.,][]{pascucci2016}. This relation was recovered by dust evolution models that include the  growth, fragmentation, and drift of particles from \cite{krijt2016}. Drift is crucial for reproducing the steepness of the  $M_{\rm{dust}}-M_{\star}$ relation because it is expected to reduce the dust mass with time and because drift is more effective around low-mass stars \citep{pinilla2013}. However, in these models, the disks are short in solids compared with observations taken after $\lesssim$1\,Myr of evolution. When considering disks with large inner cavities resolved at millimeter-wavelengths (the so-called transition disks; TDs), \cite{pinilla2018} find that this relation is much flatter, suggesting that the flatness is due to an effective reduction of radial drift in pressure bumps. 

In this paper, we aim to study the effect that dust evolution has on the $M_{\rm{dust}}-M_{\star}$ relation. We are particularly interested in investigating the effect of particle traps in the disk on this relation. Motivated by the fact that exoplanet statistics show a dependence on stellar mass and metallicity, we focus also on the effect of having different stellar and disk masses in the dust evolution models, assuming that the disk dust mass is a proxy of stellar metallicity. Furthermore, we investigate the effect of having pressure bumps of different amplitudes. In the context of planets being responsible for those pressure bumps, it is an open question of whether the $M_{\rm{dust}}-M_{\star}$ relations are, perhaps, tracing the presence of giant planets (hence stronger pressure bumps) around more massive stars, while sub-Neptune planets (hence weaker pressure bumps) around low-mass stars, in agreement with exoplanet statistics. In this picture, planets have already formed at the very first million years and the millimeter disk mass is the leftover mass, which does not drift inward and can be still observed by ALMA. 

This paper is organized as follows. In Sect.~\ref{sect:models}, we explain the dust evolution models and the assumptions made when comparing them with observations. In Sect.~\ref{sect:results}, we present the results of these dust evolution models and the comparison with current millimeter observations of planet-forming disks, in particular in the context of the  $M_{\rm{dust}}-M_{\star}$ relation. In Sect.~\ref{sect:discussion}, we discuss our results and assumptions, and the main conclusions are given in Sect.~\ref{sect:conclusion}.

\section{Methods}                       \label{sect:models}

\subsection{Dust evolution models}

To study the effect of pressure traps in the observed $M_{\rm{dust}}-M_{\star}$ relation, we performed dust evolution models that include the growth and dynamics of particles, based on the models from \cite{birnstiel2010b}. 

In these models, all the particles are initially micron-sized grains distributed as the gas with a dust-to-gas ratio of 1/100. We assumed a maximum grain size of 2\,m, and the grain size grid was logarithmically spaced in 180 cells. For the growth process, we took into account the coagulation, fragmentation, and erosion of particles; while for the dynamics, particles are influenced by the gas drag and, hence, the radial and azimuthal drift, the turbulent mixing, vertical settling, and Brownian motion. Particles stick and grow when their relative velocities before collisions are below a threshold, which is set in the simulations to be 10\,m\,s$^{-1}$. This value is based on numerical and laboratory experiments of icy particles, that show that above these velocities particles are expected to fragment \citep[e.g.,][]{blum2000, blum2018, paszun2009, gundlach2015}, although recent laboratory experiments show that ice particles may be as weak as silicate particles \citep[e.g.,][]{gundlach2018}, with fragmentation velocities of $\sim$1\,m\,s$^{-1}$. The dust diffusion is assumed to be as the turbulent gas viscosity \citep{youdin2007} and the turbulent velocities are proportional to $\sqrt{\alpha}$, which is the effective viscosity parameter for the disk evolution \citep{shakura1973}. In the models, this parameter $\alpha$ has  an influence, therefore, on the relative velocity of particles (and their fragmentation) due to turbulence and settling, as well as on the diffusion or mixing of grains.

For the models, we assumed different stellar parameters and disk masses and fixed other disk parameters. This approach is aimed at achieving a better understanding of the effect of stellar and disk mass on the radial drift and trapping of particles and, therefore, on the final $M_{\rm{dust}}$ that would be observed. Because radial drift is more efficient around low-mass stars \citep{pinilla2013, zhu2018}, we expect to obtain dust masses that decrease with decreasing stellar mass. To visualize this, we consider four different type of stars: (1) a very low-mass star with a mass of $M_\star=0.1\,M_\odot$, a luminosity of $L_\star=8\times10^{-2}\,L_\odot$, and an effective temperature of 3,000\,K. These properties are motivated by the CIDA\,1, a low-mass star whose disk has a large inner cavity resolved at millimeter wavelengths \citep{pinilla2018}; (2) an intermediate-mass star with $M_\star=0.3\,M_\odot$, with a luminosity of $L_\star=0.2\,L_\odot$, and an effective temperature of 3,700\,Kl (3) a solar-type star, that is, with $M_\star=1\,M_\odot$, $L_\star=2\,L_\odot$, and an effective temperature of 4,700\,K. And finally, (4) a Herbig star with $M_\star=2.0\,M_\odot$,  a luminosity of $L_\star=20\,L_\odot$, and an effective temperature of 9000\,K. In the first set of simulations, we keep the disk mass the same independently of the the stellar mass, such that $M_{\rm{disk}}=0.05\,M_\odot$. These models aimed to investigate the effect of radial drift around different stellar masses only. In the second set of simulations, we assumed $M_{\rm{disk}}=0.05\,M_\star$. According to protostellar disks formed in radiation hydrodynamical simulations of star-cluster formation, the disk mass is expected to be around 0.01-0.08\,$M_\star$ for ages higher than 10000 years \citep{bate2018}

\begin{figure}
 \centering
        \includegraphics[width=\columnwidth]{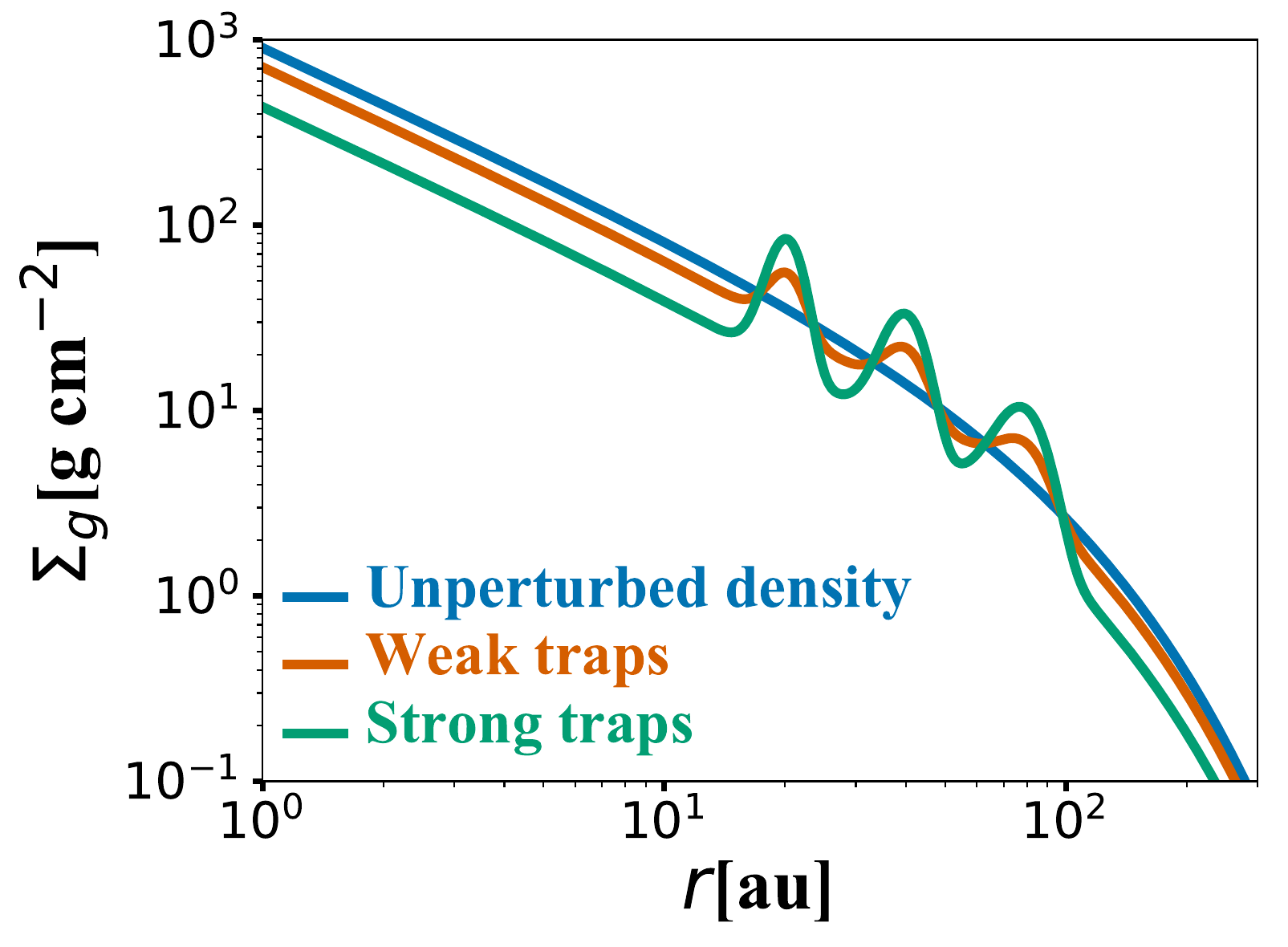}
   \caption{Assumed profiles for the gas surface density distribution when the disk mass is assumed to be $0.05\,M_\odot$ for all cases.}
   \label{initial_gas_profile}
\end{figure}

The temperature of the dust is assumed to be a power law that depends on the stellar luminosity, such that,

\begin{equation}
        T_d(r)=T_{10}\left(\frac{r}{\mathrm{10\ au}}\right)^{-1/2}\left(\frac{L_\star}{L_\odot}\right)^{1/4}
  \label{T_dust}
,\end{equation}

\noindent where $T_{10}$ is taken to be 30\,K \citep{andrews2013, tripathi2017} and $L_\star$ is the stellar luminosity. Equation~\ref{T_dust} is expected if the dust disk size is independent of stellar mass \citep{hendler2017}. 

Since the disk temperature is a function of stellar luminosity, we expect significant differences in some physical quantities across the range of stellar masses that we study. First, at a given distance, the sound speed is higher around more massive stars, making the disk scale height also higher ($h(r)=c_s\Omega^{-1}$, with $c_s$ being the sound speed and $\Omega$ the Keplerian frequency). Second, in dust evolution models, the relative velocities between particles are dominated by turbulence and radial drift. A change in temperature affects the maximum value of turbulent velocities, which are directly proportional to the sound speed, $v_{\rm{turb}}\propto\sqrt{\alpha_{\rm{turb}}}~c_s$ \citep{ormel2007}.

\begin{figure*}
 \centering
  	\includegraphics[width=18cm]{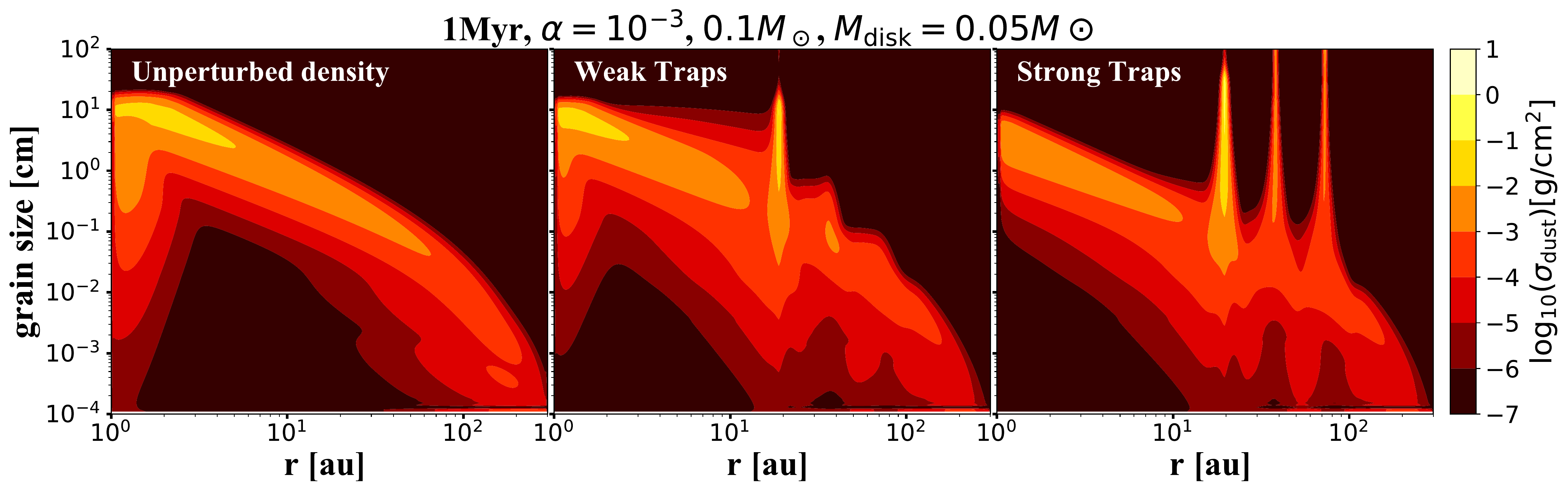}\\
  	\includegraphics[width=18cm]{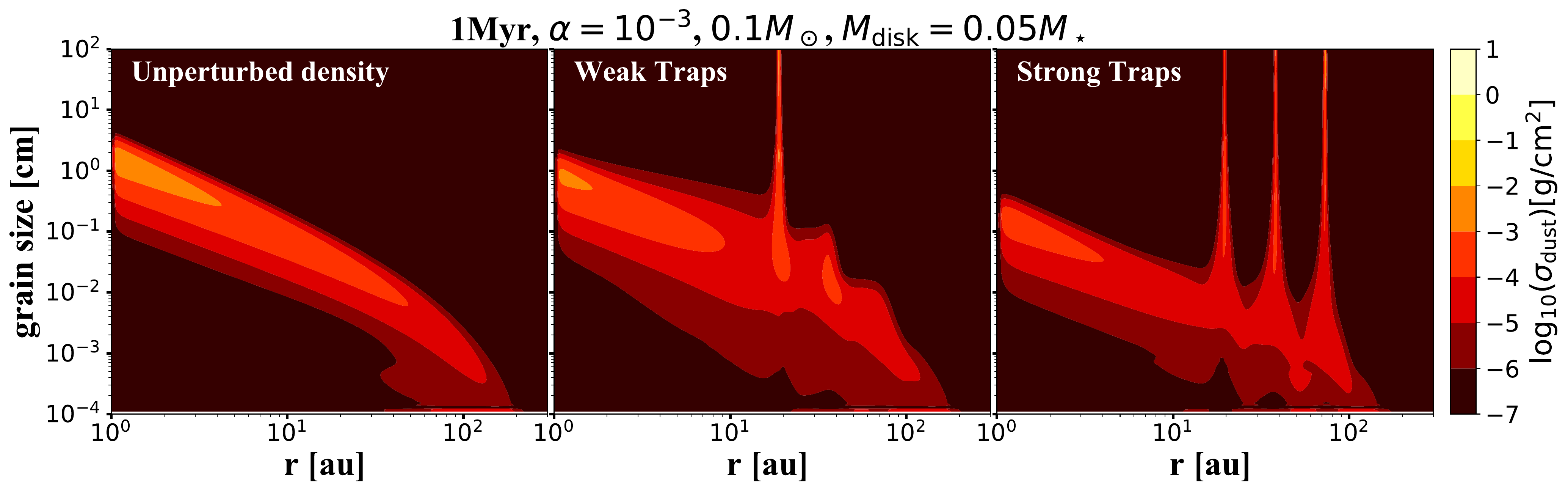}
   \caption{Dust density distribution after 1\,Myr of evolution for no traps (left panel), weak traps (middle panel), and strong traps (right panel), for the case of $M_\star=0.1M_\odot$ and $\alpha=10^{-3}$. In the top panels, the disk mass is $M_{\rm{disk}}=0.05\,M_\odot$, while for the bottom panels, the disk mass is $M_{\rm{disk}}=0.05\,M_\star$.}
   \label{dust_density_distribution}
\end{figure*}

The disk surface density distribution is parametrized by an exponentially tapered power-law function, given by

\begin{equation}
        \Sigma_{\rm{gas}}(r)=\Sigma_0\left(\frac{r}{R_c}\right)^{-\gamma} \exp\left[-\left(\frac{r}{Rc}\right)^{2-\gamma}\right]
  \label{Sigma_disk}
,\end{equation}

\noindent where $\gamma=1$, $R_c=80\,$au, and $\Sigma_0$ is taken such that the disk mass is either $0.05\,M_\odot$ or $0.05\,M_\star$ . The radial grid is taken from 1 to 300\,au and it is logarithmically spaced in 300 cells.

\paragraph{Particle traps.} To include the pressure bumps in the disk, we assume Gaussian perturbations to the disk surface density given by

\begin{equation}
        B(r)=A\exp\left(-\frac{(r-r_p)^2}{2w^2}\right)
  \label{Gaussian_perturbation}
,\end{equation}

\noindent where $A$ is the amplitude, $r_p$ the center, and $w$ the width of the Gaussian perturbation. We assume three pressure bumps in the disk motivated by the average amount of sub-structures observed in protoplanetary disks at high angular resolution \citep{huang2018, long2018} and claimed to be particle traps \citep{dullemond2018}. Hence, the gas surface density of the perturbed disk is given by

\begin{equation}
        \Sigma^\prime_{\rm{gas}}(r)=\Sigma_{\rm{gas}}(r)\times[1+B_1+B_2+B_3]
  \label{Sigma_disk_perturbated}
,\end{equation}

\noindent where $B_1$, $B_2$, and $B_3$ are given by Eq.~\ref{Gaussian_perturbation}. We consider two different types of perturbations. One type we refer to as strong pressure bumps, with $A=4$; and another that we call weak pressure bumps, with $A=1$. This simple parametric form aims to mimic the presence of multiple giant planets or (sub-) Neptune planets creating pressure bumps exterior to their orbits. We note that, in reality, $A$ depends not only on the planet mass, but also on several disk parameters, such as viscosity and local scale height  \citep[e.g.,][]{crida2006, fung2014}. When comparing the assumed gas surface density with the hydrodynamical simulations from \cite{zhang2018},  who studied three different disk scale heights and three different values of $\alpha$  for 4 different planet masses, the amplitude of our perturbations resemble planet masses of 33\,$M_{\rm{Earth}}$ and 0.3\,$M_{\rm{Jup}}$ in the case of $\alpha=10^{-3}$ and $h/r=0.05$ when $A=4$ and $A=1$, respectively; or 0.3\,$M_{\rm{Jup}}$ and 1\,$M_{\rm{Jup}}$ when $h=0.07-0.1$ (for $A=4$ and $A=1$, respectively). For this $\alpha$ value and planet masses, the gap width does not change significantly with the planet mass, so our assumption of keeping the same width independently of $A$ is fair for this range of planet masses and disk scale heights. 

\begin{table}
\caption{Parameters of the model}    
\label{model_parameters}     
\centering                         
\begin{tabular}{c c }       
\hline\hline                 
Parameter & Values \\    
\hline                       
   $A$ & $1, 4$  \\ 
   $r_p$ [au] & $20,40,80$  \\ 
   $w$ [au] & $2,5,11$  \\                                             
   $\alpha$ & $10^{-4},10^{-3},10^{-2}$  \\   
   $M_{\rm{disk}}$ & $0.05\,M_\odot$ or $0.05\,M_\star$  \\
   $R_c$ [au] & $80$  \\
   $\gamma$ & $1$  \\
   $M_\star [M_\odot]$ & $0.1, 0.3, 1.0, 2.0$  \\
   $v_f$[m\,s$^{-1}$]& $10$  \\
\hline                     
\end{tabular}  
\end{table}

In both cases, the Gaussian perturbations are centered at 20, 40, and 80\,au; and the width is the local scale height $h(r)$. Because we assumed different stellar masses and luminosities (and, hence, different disk temperatures), the local scale height is different for each case. However, we set the width of the perturbation as the disk scale height that corresponds to the case of a very low mass star; this means that for all cases, $w$ has values of $\sim$2, $\sim$5, and, $\sim$11\,au at 20, 40, and 80\,au, respectively. The width of the perturbation is, therefore, larger than the local scale height for the other three stellar masses, becoming almost a factor of two for the Herbig disk. Although the Herbig disk is warmer and, hence, it is expected that the disk scale height is higher at a given location, the dependence on the stellar mass makes that the disk scale height increases for the low mass stars. Hence, the widths are larger or equal to the pressure scale height for all cases to ensure the stability of the pressure bumps \citep[e.g.,][]{pinilla2012, dullemond2018}. The width of the perturbation should not be confused with the width of the dust rings (see e.g., DSHARP) as the latter can be much smaller than the width of the pressure bump and of the local scale height.

\begin{figure*}
 \centering
    \includegraphics[width=18cm]{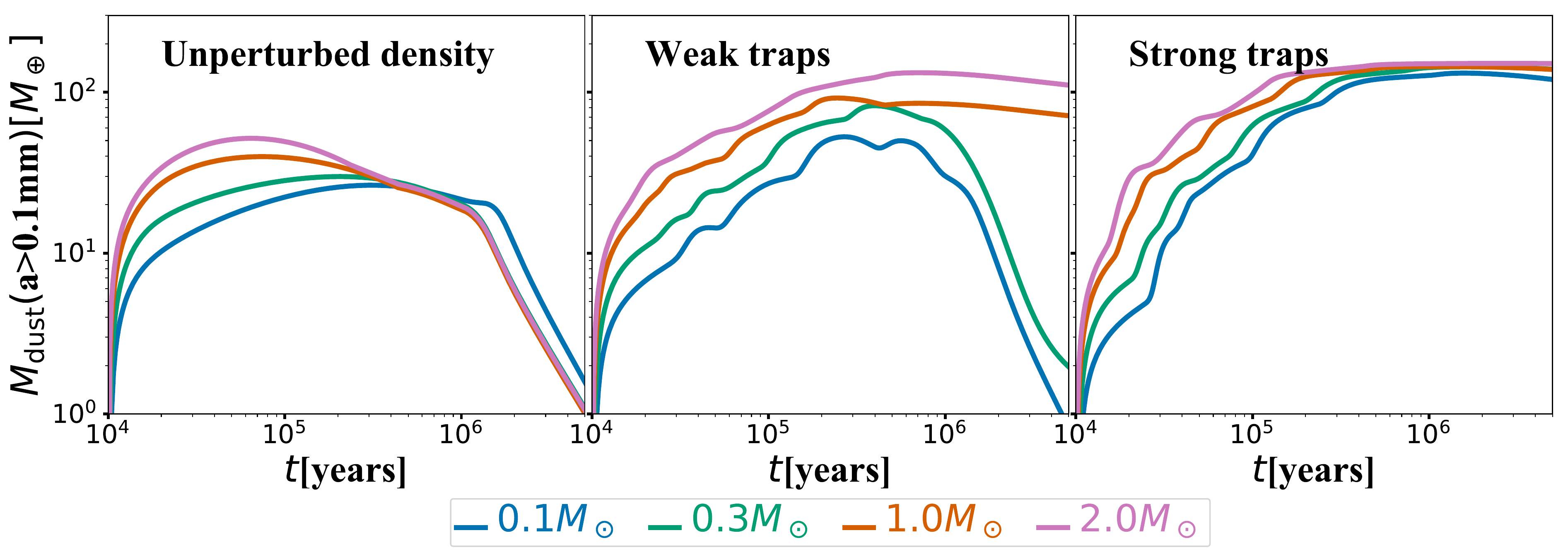}
   \caption{Effect of stellar mass on the evolution of dust mass of particles larger than 0.1 mm from the models with no traps (left panel), weak traps (middle panel), and strong traps (right panel). In all cases $\alpha=10^{-3}$ and   $M_{\rm{disk}}=0.05\,M_\odot$. }
   \label{Mstar_effect}
\end{figure*}

For the two types of pressure bumps, the perturbed gas surface density is normalized such that the disk mass remains as the unperturbed cases, that is, 0.05\,$M_\odot$ or 0.05\,$M_\star$. Figure~\ref{initial_gas_profile} shows the gas surface density profile for the case without any traps (unperturbed density) and the cases with strong and weak pressure traps when  $M_{\rm{disk}}$=0.05\,$M_\odot$. Table~\ref{model_parameters} summarizes the model parameters considered in this work. 

\subsection{Synthetic dust mass} \label{sect:mock_obs}

In order to compare the results of the models with observations, we took the dust density distribution at different times of evolution and calculated the vertical optical depth ($\tau_\nu$) at 0.87\,mm, which corresponds to ALMA Band\,7, the band adopted for most of the medium-resolution surveys of  star-forming regions.

To calculate $\tau_\nu$, we took the  vertically integrated dust density distribution $\sigma(r,a)$ at a given time of evolution and calculate $\tau_\nu=\sigma(r,a) \kappa_\nu/\cos i$, where $\kappa_\nu$ is the opacity for each grain size and at a given frequency or wavelength ($\kappa_\nu$). To obtain $\kappa_\nu$, we use Mie theory and we assume the same volume fractions and optical properties for the dust particles as in \cite{ricci2010}. Furthermore, we consider face-on disks only, that is, $i=0$. This assumption has no effects for optically thin discs, but it can lead to overall higher disc fluxes and thus higher predicted masses if discs are close to being optically thick.
With the optical depth, the intensity profile at a given wavelength is given by
\begin{equation}
        I(r)=B_\nu (T (r))\left[1-e^{-\tau_\nu(r)}\right].
  \label{intensity}
\end{equation}

We obtain the total flux as

\begin{equation}
        F_\nu=\frac{2\pi\cos{i}}{d^2}\int_{r_{\mathrm{in}}}^{r_{\mathrm{out}}} I(r) r dr,
  \label{eq:flux}
\end{equation}

\noindent where $d$ is the distance to the source, which we set to 140\,pc, and $B_\lambda(T(r))$ is the Planck function, for which we take the temperature profile of Eq.~\ref{T_dust}.  

To obtain the dust mass as is usually calculated from the millimeter fluxes, we assume optically thin emission \citep{hildebrand1983} and, hence,

\begin{equation}
        M_{\mathrm{dust}}\simeq\frac{{d^2 F_\nu}}{\kappa_\nu B_\nu (T(r))}.
  \label{mm_dust_mass}
\end{equation}

For  \ $\kappa_\nu$, we assume what is usually taken in disk surveys, which is a frequency-dependent relation given by  $\kappa_\nu=2.3\,$cm$^{2}$\,g$^{-1}\times(\nu/230\,\rm{GHz})^{0.4}$ \citep[][]{andrews2013}, and for $B_\nu$ we assume a blackbody surface brightness corresponding to 20\,K. 

\section{Results}                       \label{sect:results}

\subsection{Stellar mass effect}

In this section, we first describe the results when $M_{\rm{disk}}=0.05\,M_\odot$ for all cases and focus on the effect of having different stellar mass. The top panels of Fig.~\ref{dust_density_distribution} show the dust density distribution after 1\,Myr of evolution and $\alpha=10^{-3}$ for the cases of no traps (left panel), weak traps (middle panel), and strong traps (right panel) for a stellar mass of $0.1\,M_\odot$.  This figure shows the effect of particle trapping in pressure bumps and the effect of their amplitude.  

\paragraph{Unperturbed density.} In the absence of particle traps, grains grow from micron-sized particles to the maximum grain size that is seen in the case of $0.1\,M_\odot$ set by fragmentation in most of the disk. When particles reach centimeter or millimeter sizes, they drift towards the star, depleting the disk of pebbles in million-year timescales. This is shown in the left panel of Fig.~\ref{Mstar_effect}, which illustrates the effect of stellar mass on the evolution of the total mass in particles larger than 0.1\,mm (hereafter, called mass in large dust). In this case, the maximum mass in large dust is quickly reached after few thousand of years of evolution and decreases with time until few large dust particles remain in the disk. The maximum of the total mass in large dust is around $50\,M_\oplus$ (30\% of the initial mass in solids) for the Herbig star and it is around $25\,M_\oplus$ for the low-mass star, even when in all cases, the disk (and dust) mass was initially the same. This difference is a result of the growth and more efficient drift around low-mass stars. However, after $\sim$1\,Myr of evolution, the mass in large dust is almost the same for the four stellar masses ($\sim20\,M_\oplus$). At 5\,Myr of evolution, the dust mass has decreased to less than 1\,$M_\oplus$. 

\paragraph{Particle traps.} When particle traps are included in the disk, particles can grow to millimeter or centimeter sizes and remain trapped in pressure maxima (middle and right panels of Fig.~\ref{dust_density_distribution} and Fig.~\ref{Mstar_effect}). In the case of weak traps (with A=1.0 in Eq.~\ref{Gaussian_perturbation}), the amplitude of the pressure bumps is enough to keep a high amount of large dust in the disk for the cases of a Solar-type star and a Herbig star (2\,$M_\odot$). The mass in large dust is around 130\,$M_\oplus$ and 110\,$M_\oplus$ for the 2\,$M_\odot$ stellar mass at 1 and 5\,Myr of evolution, respectively (Fig.~\ref{Mstar_effect}). For the Solar type star the mass in large dust is around 85\,$M_\oplus$ and 70\,$M_\oplus$ at 1 and 5\,Myr of evolution, respectively. For the 0.1\,$M_\odot$ and 0.3\,$M_\odot$ stars, these pressure bumps are not enough to retain millimeter- or centimeter-sized particles and after around $0.5\,$Myr of evolution the mass in large dust decreases with time reaching values below 2\,$M_\oplus$ at 5\,Myr of evolution. This is because of the more efficient radial drift around low-mass stars \citep{pinilla2013}. The radial drift of particles is a consequence of the difference between the gas azimuthal velocity  and the Keplerian speed, which is higher for disks around low-mass stars. Therefore, the weak traps cannot efficiently trap particles for the two low-mass stars that we consider (middle panel of Fig.~\ref{Mstar_effect}).

When strong pressure bumps are assumed in the dust evolution models, their amplitude is high enough to retain most of the large dust independent of the stellar mass at million-year timescales (right panels of  Figs.~\ref{dust_density_distribution} and \ref{Mstar_effect}). From 1\,Myr of evolution, the mass in large dust range between 150 to 120\,$\,M_\oplus$ for the most massive star (2\,$M_\odot$) and the lowest mass star (0.1\,$M_\odot$), respectively. When trapping is effective, increasing the amplitude does not change significantly the mass in large dust with time. This can be seen when comparing the results from weak and strong traps, where the mass in large dust in the cases of disks around 1 and 2\,$M_\odot$ is very similar independently of the amplitude of the pressure bumps (middle and right panel of Fig.~\ref{Mstar_effect}). Therefore, with an amplitude of $A=4$, we  reach a saturation in trapping for all the stellar masses \citep[see also Fig.\,6 in ][]{pinilla2012}.

Due to the accumulation of dust particles in these strong pressure maxima, the disk emission becomes optically thick in several disk locations. In addition, dust grows as large as the maximum grain size that we allow in the simulations (right panels of  Fig.~\ref{dust_density_distribution}). Both effects result on an underestimation of the dust disk masses when they are calculated from sub-millimeter fluxes, as discussed in Sect.~\ref{sect:obs}. 

\begin{figure}
 \centering
        \includegraphics[width=\columnwidth]{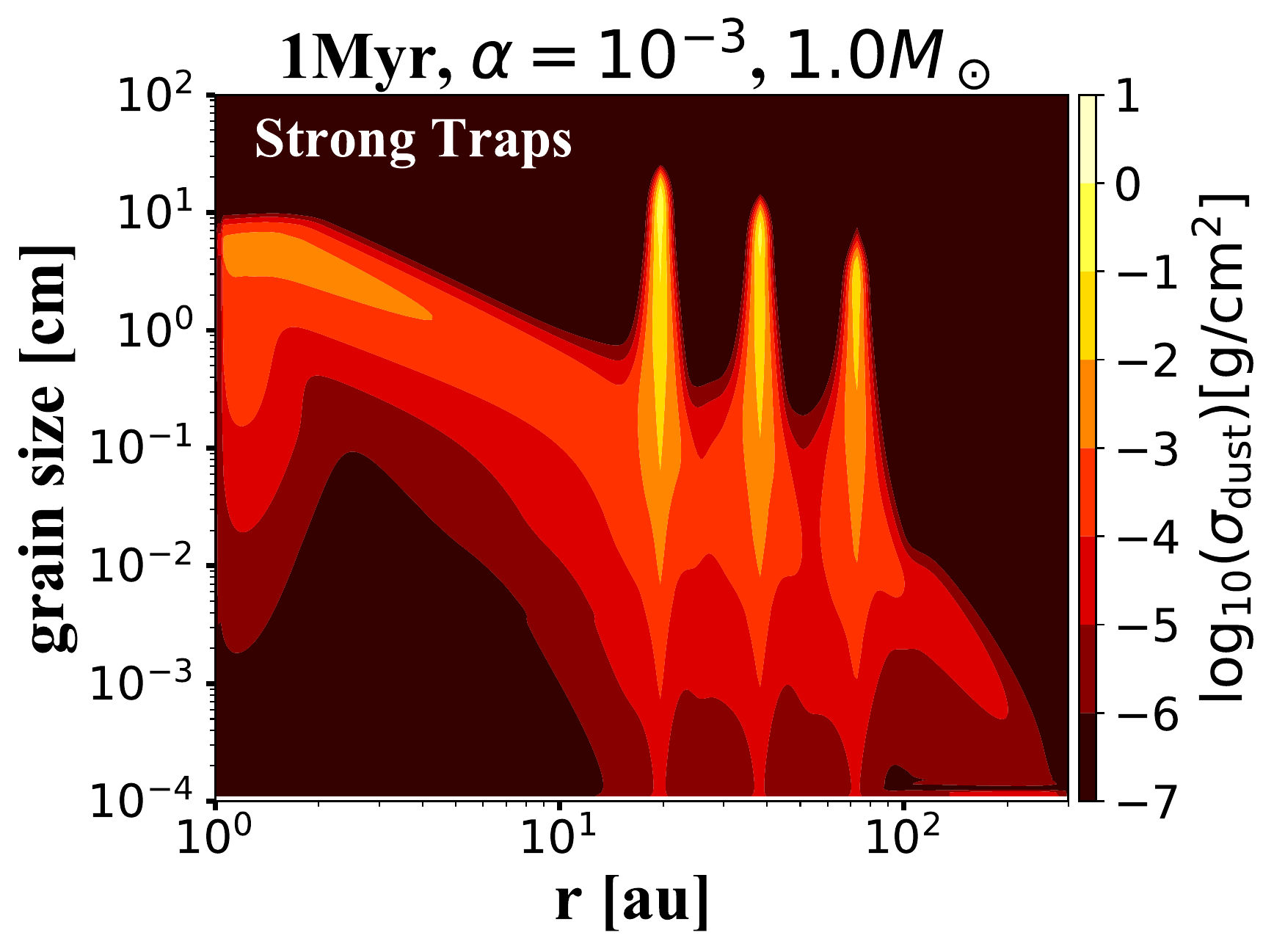}
   \caption{Dust density distribution after 1\,Myr of evolution for the case of strong traps, $M_\star=1.0M_\odot$, $M_{\rm{disk}}=0.05\,M_\odot$, and $\alpha=10^{-3}$.}
   \label{dust_density_distribution_1Msun}
\end{figure}

\paragraph{Effect of the stellar mass on the fragmentation barrier.} The maximum size that particles can reach before they fragment, or the fragmentation  barrier, is given by \citep{birnstiel2012}:

\begin{equation}
        a_{\rm{frag}}=\frac{2}{3\pi}\frac{\Sigma}{\rho_s \alpha_{\rm{turb}}}\frac{v_{\rm{frag}}^2}{c_s^2}.
  \label{afrag}
\end{equation}

When the disk surface density and $\alpha-$viscosity are the same in the simulations, the effect of changing the stellar mass and luminosity is reflected in the sound speed. For a more luminous star, that is, with a warmer disk, the sound speed is higher, increasing the maximum turbulent velocity and therefore decreasing $a_{\rm{frag}}$. As a result, in a warmer disk, particles reach lower sizes. Figure~\ref{dust_density_distribution_1Msun} shows the case of strong pressure bumps in the case of a disk around a Solar type star. The maximum grain size that particles can reach, is lower than in the case of a $0.1\,M_\odot$ stellar mass (top right panel of Fig.~\ref{dust_density_distribution}), an effect that is more clear inside the pressure bumps. In the case of a Solar-type star, more millimeter or centimeter grains remain in the disk at million-year timescales because fragmentation keeps the particles smaller and closer to the sizes where they are more affected by the gas drag, which means that they are easier to trap in pressure bumps. This effect will influence the synthetic sub-millimeter fluxes that are derived from the models because when growth is very efficient (as in the case of strong bumps and low-mass stars), the very large grains ($\gtrsim10$\,cm) that are formed inside the pressure maxima have very low opacities contributing very little to the millimeter fluxes.

\begin{figure}
 \centering
        \includegraphics[width=\columnwidth]{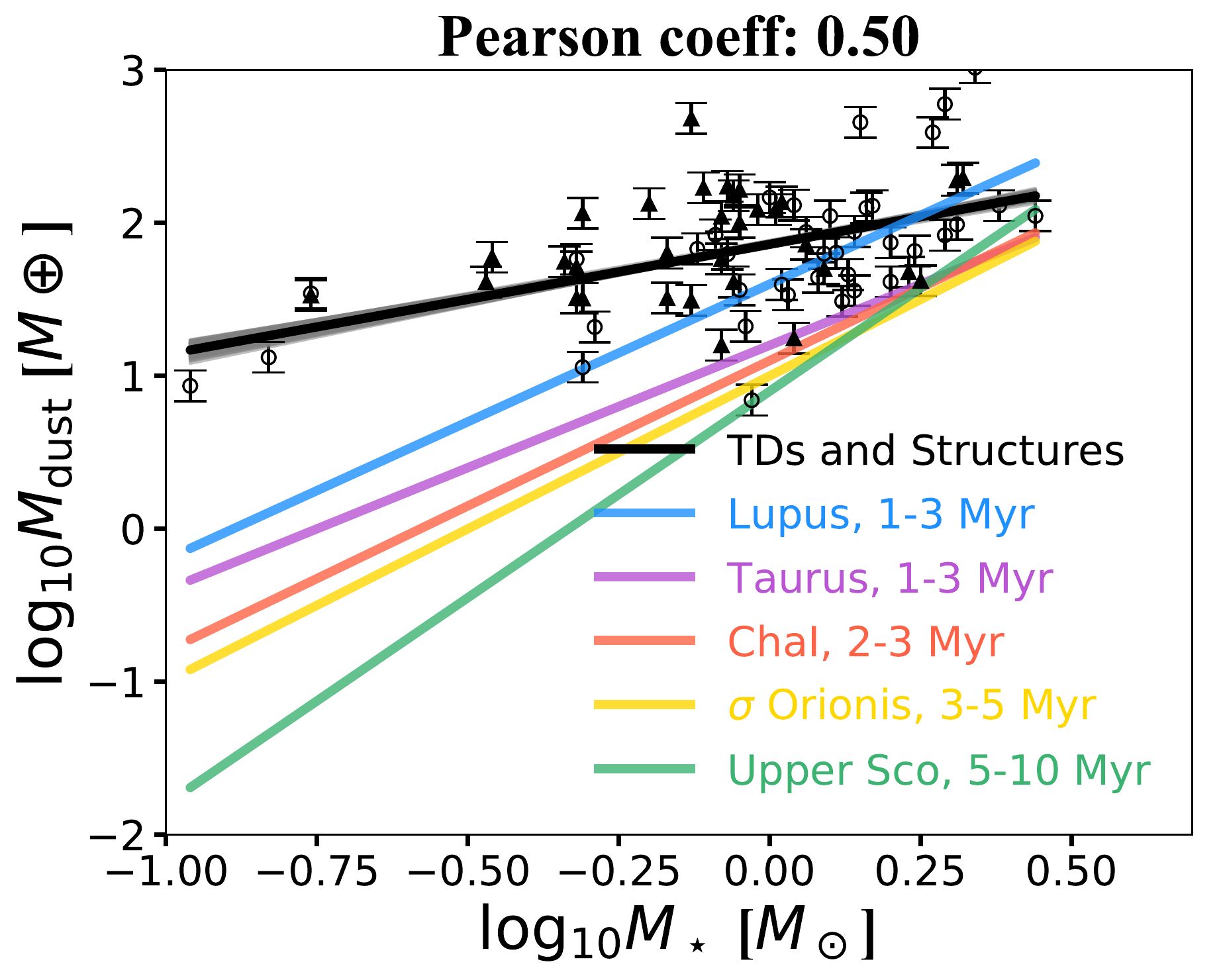}
   \caption{$M_{\rm{dust}}-M_{\star}$ relation in different star-forming regions (colors) and for the disks with resolved cavities or transition disks  (TDs, open circles) and disks with structures (full triangles, see Appendix). The values for the slope and intercept for the different star forming regions are taken from \cite{pascucci2016}, except for $\sigma$-Orionis, which values are  from \cite{ansdell2017}. From fitting this relation for TDs and disk with structures, we used $\log_{10}(M_{\rm{dust}}/M_\oplus)=\beta \log_{10}(M_\star/M_\odot)+\alpha$, and we obtained $\beta=0.72^{+0.04}_{-0.04}$  and $\alpha=1.86^{+0.01}_{-0.01}$ and a Pearson coefficient of 0.50. The fit takes into account the uncertainties of the data, which are dominated by the 10\% of uncertainty from flux calibration.}
   \label{Mdisk_Mstar_relation}
\end{figure}

\subsection{Disk mass effect}

There are two main effects of changing the disk mass to be a fraction of the stellar mass (0.05\,$M_\star$) instead of assuming  0.05\,$M_\odot$ for all the cases. First of all, the fragmentation barrier ($a_{\rm{frag}}$, Eq.~\ref{afrag}) is directly proportional to the gas surface density ($\Sigma$), implying that $a_{\rm{frag}}$ is lower in low-mass disks even though the sounds speeds are lower. On the other hand,  the radial drift of particles comes from the gas drag forces, which depend on dust properties such as their mass and cross-sectional area, but also on the gas disk via the gas density. The dimensionless Stokes number (St) measures the importance of drag force by comparing the stopping time of the particles to the Keplerian frequency at a given radius. Particles with the Stokes number equal unity are subject to the highest radial drift. At the midplane, the Stokes number is inversely proportional to the gas surface density and as a consequence, a millimeter-sized particle in the outer disk has a higher Stokes number in a lower mass disk. Therefore, the drift barrier is also lower in lower mass disks, which means that particles grow to smaller sizes in a low-mass disk, but also a particle  with a given size drifts faster in a low-mass disk. 

\begin{figure*}
 \centering
    \includegraphics[width=18cm]{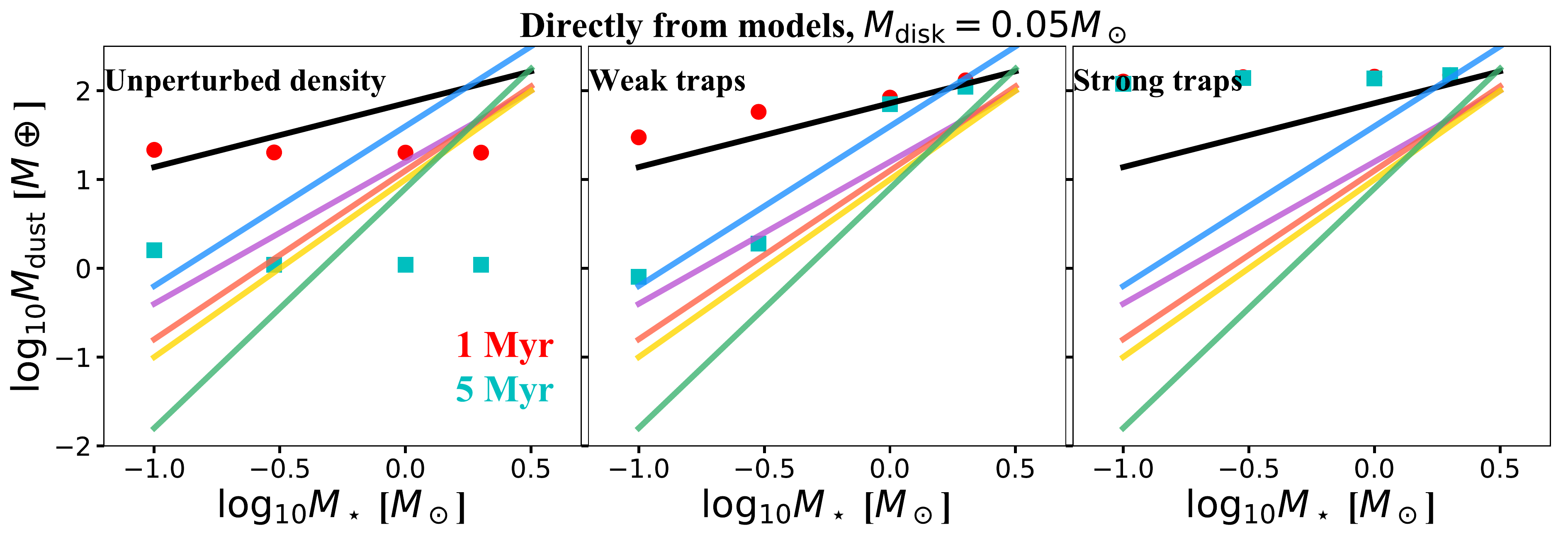}\\
    \includegraphics[width=18cm]{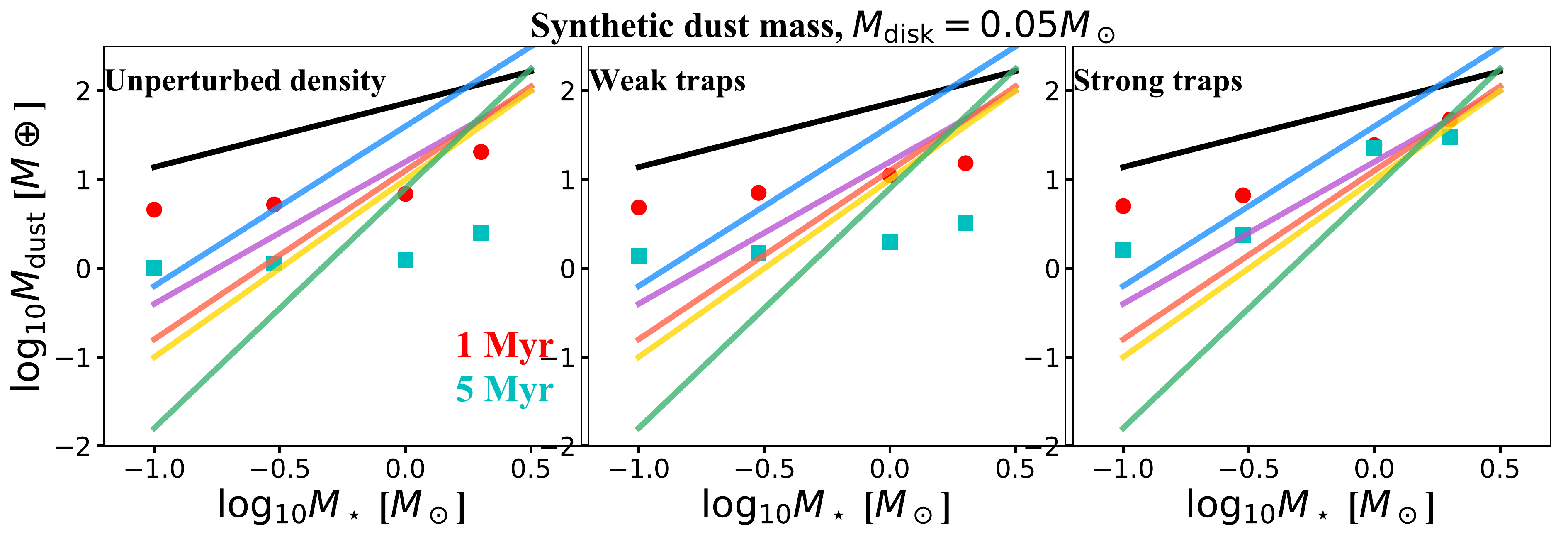}
   \caption{Comparison with observations, models with $\alpha=10^{-3}$ and $M_{\rm{disk}}=0.05\,M_\odot$ in all cases. Top panels: total mass of the large dust directly obtained from models. Bottom panels: synthetic dust mass after considering assumptions from observations.}
   \label{Mstar_Mdisks_models}
\end{figure*}

Figure~\ref{dust_density_distribution} compares the case of 0.1\,$M_\odot$ stellar mass when  $M_{\rm{disk}}=0.05\,M_\odot$ vs. $M_{\rm{disk}}=0.05\,M_\star$ and shows how for the case when $M_{\rm{disk}}=0.05\,M_\star$, the maximum grain size is lower and due to more efficient drift of millimeter- or centimeter-sized particles, the disk is more depleted of these kind of grains. As result, in the cases with traps, there is less effective trapping and fewer millimeter grains remain inside pressure maxima.

\subsection{Comparison with the observations: $M_{\rm{dust}}-M_{\star}$ relation} \label{sect:obs}

In order to compare the results from dust evolution models with current millimeter observations of protoplanetary disks, in particular in the context of the  $M_{\rm{dust}}-M_{\star}$ relation, we used the values for the slope and intercept for different star-forming regions from \cite{pascucci2016} as stellar masses were homogeneously derived in this study, except for $\sigma$-Orionis, which was not included; hence, for this parameter, we took the values from \cite{ansdell2017}. In all regions, $T=20\,$K was assumed (see solid lines in different colors in Fig.~\ref{Mdisk_Mstar_relation}). We also collected from the literature a sample of disks that have been observed with ALMA, which have either a large inner cavity or gap, defined as transition disks (TDs, empty circles), and disks with sub-structures, such as rings, gaps, or asymmetries, but still exhibiting substantial emission from the inner disk (full triangles). We selected the TDs that have been observed with ALMA and whose cavities are resolved, which implies that the TDs in our sample have large cavities ($\gtrsim$20\,au). We gathered a total of 73 disks (43 TDs and 30 disks with substructures, see Table in the appendix). The sample of TDs double the number of targets presented previously in \cite{pinilla2018}. 

We placed the TDs and disks with substructures in the  $M_{\rm{dust}}-M_{\star}$ relation and fit these points with the following relation: $\log_{10}(M_{\rm{dust}}/M_\oplus)=\beta \log_{10}(M_\star/M_\odot)+\alpha$. We obtained $\beta=0.72^{+0.04}_{-0.04}$  and $\alpha=1.86^{+0.01}_{-0.01}$, and a Pearson coefficient of 0.50. The fit takes into account the uncertainties of the data, dominated by the 10\% of uncertainty from flux calibration. When only the TDs are considered for the fit, the Pearson coefficient is higher (0.67) and, therefore, this correlation is stronger when only TDs are included. This is because for this correlation, we mainly have data for TDs around low-mass stars. If only disks with substructures are considered, the correlation is weak with a Pearson coefficient of 0.28. However, more observations in the low stellar mass regime that include the resolution of substructures are needed to confirm if there is indeed a correlation. 

The values for the slope and intercept change when including or excluding disks with substructures. When we only consider TDs, the values remain similar with $\beta_{\rm{TD}}=0.94^{+0.06}_{-0.06}$  and $\alpha_{\rm{TD}}=1.80^{+0.02}_{-0.02}$, but the relation is shallower when only disks with substructures are considered, with $\beta_{\rm{SB}}=0.42^{+0.1}_{-0.1}$  and $\alpha_{\rm{SB}}=1.90^{+0.02}_{-0.02}$. With the current data, there is not any substantial evidence that the population of TDs is different than the population of disks with substructures based on the dust disk mass. To compare the two samples, we calculate the Kolmogorov-Smirnov (KS) two-sided test, and we find a maximum deviation between the cumulative distributions of only 0.15 with a high probability ($\sim$80\%) that the two samples are similar. Overall, the $M_{\rm{dust}}-M_{\star}$ correlation of the whole sample is dominated by TDs, and it is shallower than seen from all disks in nearby star-forming as found in \cite{pinilla2018}, to which some of these disks actually belong to.

\begin{figure*}
 \centering
    \includegraphics[width=18cm]{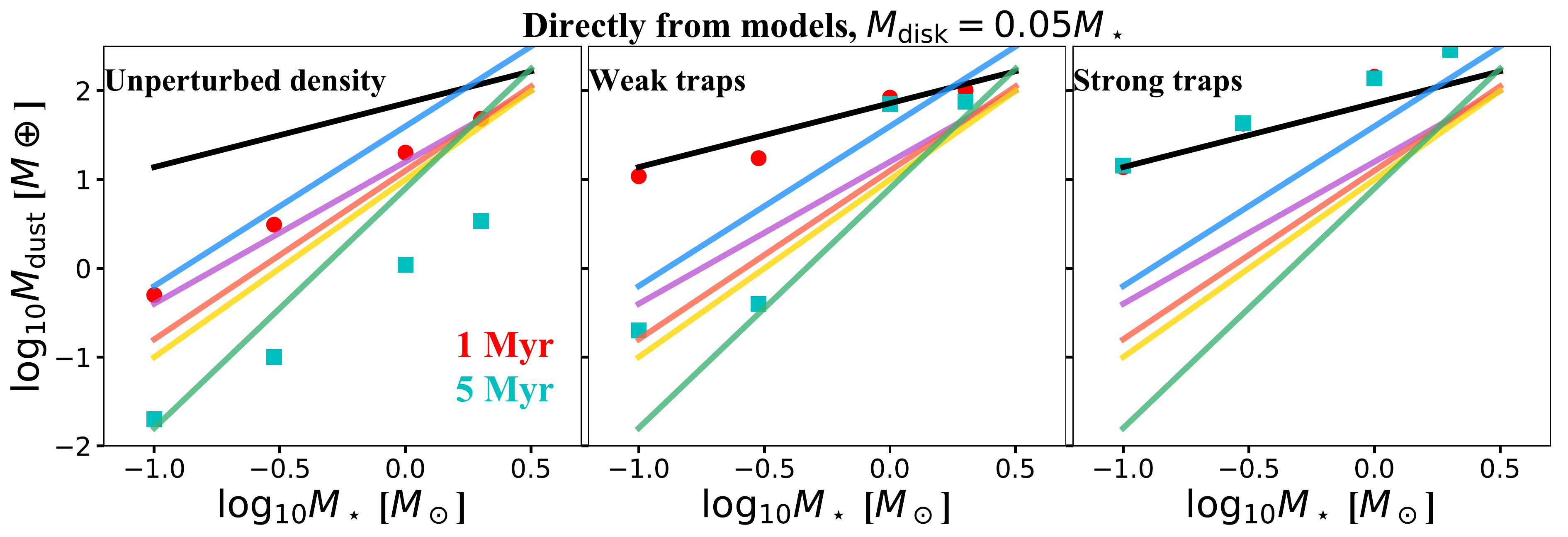}\\
    \includegraphics[width=18cm]{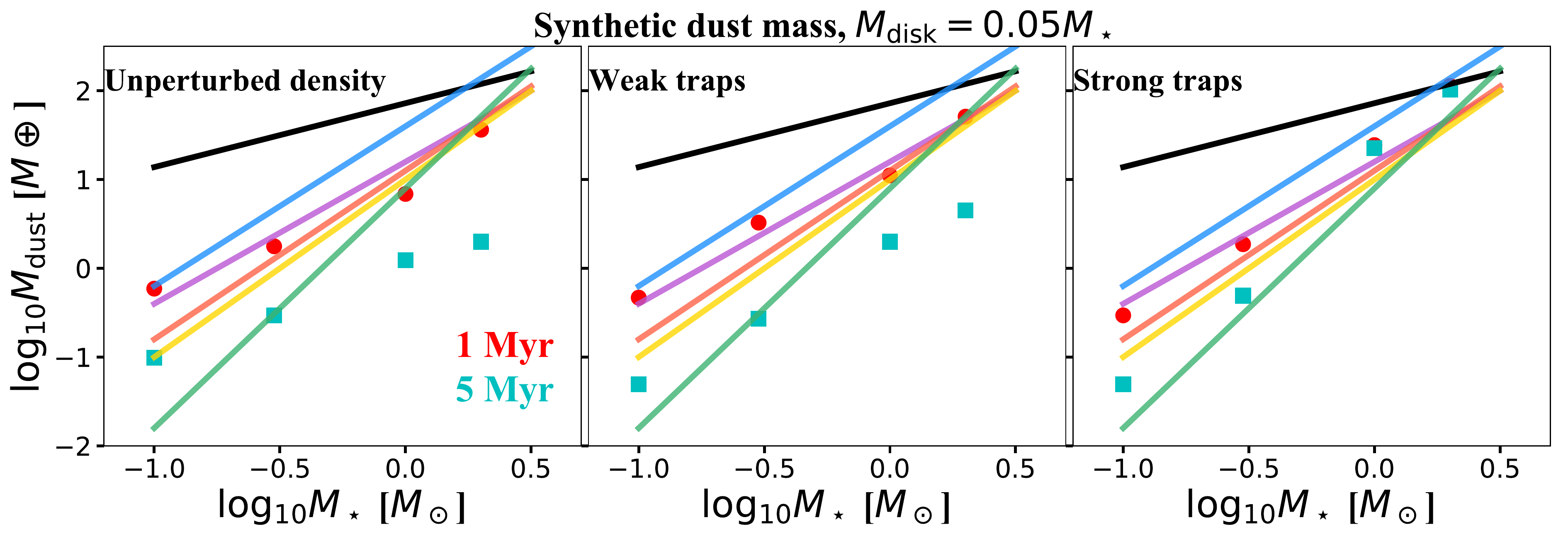}
   \caption{As in Fig~\ref{Mstar_Mdisks_models}, but with $M_{\rm{disk}}=0.05\,M_\star$ in all cases. }
   \label{Mstar_Mdisks_Mdisk}
\end{figure*}

Next, we overplotted the $M_{\rm{dust}}$ obtained from the models at 1 and 5\,Myr of evolution with the observed correlations. We plotted $M_{\rm{dust}}$ that were directly obtained from the dust evolution models for large dust without any post-processing assumption (as e.g., the values in Fig.~\ref{Mstar_effect}) and, after dust temperature and opacities were assumed, we calculated the dust mass expected from millimeter observations as explained in Sect.~\ref{sect:mock_obs} (hereafter called \emph{synthetic} dust mass). Figure~\ref{Mstar_Mdisks_models} shows the results for all the models that assumed $M_{\rm{disk}}=0.05\,M_\odot$ and Fig.~\ref{Mstar_Mdisks_Mdisk} shows the results for the cases where $M_{\rm{disk}}=0.05\,M_\star$.

\subsubsection{Synthetic dust mass compared to mass in large dust from models}
In general, the synthetic dust mass is lower than the mass in large dust that is obtained directly from the models (Fig.~\ref{Mstar_Mdisks_models} and Fig.~\ref{Mstar_Mdisks_Mdisk}). This difference is because of two different reasons. First, some parts of the disk become optically thick. In the case of an unperturbed density profile, the optical depth of the disk ($\tau$) reaches values of $\tau\sim0.1$ in the case of low-mass stars ($<M_\odot$), and $\tau\sim0.5$ in the case of higher stellar masses at 1\,Myr. The higher $\tau$ around more massive stars, even when the disk mass is assumed to be the same independently of the stellar mass,  is due to the drift more efficient removal of large particles in the case of low-mass stars. At 5\,Myr of evolution, the optical depth is much lower than unity in all the cases without particle traps, while the synthetic dust mass and the mass in large dust from the models are fairly in agreement.  

In the case of particle traps, the optical depth can reach values of between 1 to 10 inside the bumps. For the weak pressure bumps, these high values for $\tau$ are reached only in the innermost bump, while in the case of strong bumps, these values are reached at the peak of all the bumps. In the cases of particle traps, a second reason plays an important role to understand why the synthetic dust mass is lower than the mass in large dust from models, and it is due to the growth of solids beyond 10cm, hereafter referred to as boulders (in the Minimum Mass Solar Nebula (MMSN) model, the (sub-)meter-sized objects that are in between being strongly influenced by the gas and particles that are completely decoupled and moving with Keplerian speed are called "boulders".)

Boulders have very low opacities and their contribution to the millimeter fluxes is insignificant, which means that their mass is hidden when calculated from the millimeter-fluxes. This effect is clear when comparing the masses in the case of strong bumps. For low-mass stars (0.1 or 0.3\,$M_\odot$), boulders are formed inside the bumps as discussed in the previous section (Fig.~\ref{dust_density_distribution}), while in the case of disks around $1$ or $2$\,$M_\odot$, the fragmentation barrier keeps the maximum grain size around millimeter and centimeter sizes in all the pressure bumps (Fig.~\ref{dust_density_distribution_1Msun}). The formation of boulders reduces the optical depth inside the pressure bumps, but simultaneously, a lot of the solid material becomes invisible when observed at millimeter-wavelengths. As a consequence, while the synthetic dust mass is lower by a factor of $\sim1.5-2$ for the disks around stars $M_\star\geq M_\odot$, the synthetic dust mass is a factor $\sim10-100$ lower than the total mass in large dust for $M_\star<M_\odot$. Hence, in the case of traps, the difference between the synthetic dust mass and the mass from the models originates from a combination of these two effects: optical thickness and boulder formation. For the weak traps, these two effects appear only in the innermost bump at 1\,Myr of evolution. While for strong traps, these two effects appear in all the pressure traps at $\geq 1$\,Myr.

In summary, both optically thick regions and boulder formation lead to a large difference between the synthetic dust mass and the mass in large dust obtained directly from the models. Our result show that boulder formation has a more significant influence.  Recently, \cite{stammler2019} showed that to explain the  optical depth between 0.2 and 0.5 inside the rings observed in the DSHARP sample \citep{andrews2018}, boulder and planetesimal formation must occur inside the pressure bump to reduce the optical depth. In our models, we do not reduce the mass of the dust inside the pressure bumps by assuming that boulders were formed when high dust-to-gas density ratio is reached, but rather particles naturally grow to sizes beyond decimeter when drift and turbulence are inefficient to fragment or keep the particles with millimeter to centimeter size.

\subsubsection{Synthetic dust mass versus observations: $M_{\rm{disk}}=0.05\,M_\odot$ case}
In the less realistic case of disk masses independent of stellar masses, the observed slope in different star forming regions is not reproduced by the models, not when comparing directly with the models nor after post-processing to compare with the observations (see Fig.~\ref{Mstar_Mdisks_models}). Only in the case of weak traps, the slope of the sample of TDs and disks with substructures is reproduced when comparing the mass directly from the models at 1\,Myr of evolution. When comparing it with the synthetic dust mass, the slope of the TDs and disks with substructures is similar to the models with and without traps, but values are lower than observed, meaning that the initial dust mass should be higher for all stellar masses to have an agreement with the relation of TDs and substructured disks, an assumption that may be unlikely since with the assumed disk mass of $0.05\,M_\odot$, the stellar-disk mass ratio is already very high for the disks around $0.1\,M_\odot$ (stellar-disk mass ratio of 0.5) and $0.3\,M_\odot$ (stellar-disk mass ratio of 0.17). 

Strong pressure bumps are needed to produce the high disk dust masses that are observed around stars with masses equal or higher than the Solar mass (when comparing with the synthetic dust mass). With weak or no pressure bumps, the millimeter fluxes and hence the dust disk masses are lower than observed already at 1\,Myr of dust evolution.  In the context of planets causing these pressure bumps, this implies that if the disk mass is initially the same for all disks around different stellar masses, only the strong pressure bumps (or giant planets) can explain the observed values of $M_{\rm{dust}}$ for stellar masses higher than a solar mass. For low-mass stars no conclusion can be given besides that in all scenarios the synthetic dust mass is overestimated at 1\,Myr. At 5Myr, the synthetic dust mass values are in agreement with the values in Lupus, Taurus and ChaI, although the averaged age of these regions is lower than 5\,Myr.

\subsubsection{Synthetic dust mass vs. Observations: $M_{\rm{disk}}=0.05\,M_\star$ case} 
In the more realistic case of disk masses scaling with stellar masses, the unperturbed density can explain the observed $M_{\rm{dust}}-M_{\star}$ relation for different star forming regions only at 1\,Myr when compared directly with the mass in large dust obtained from the models (see Fig.~\ref{Mstar_Mdisks_Mdisk}). When comparing with the synthetic dust mass predicted by our model, the slope at 1\,Myr is slightly lower than in the actual observations. We find $M_{\rm{dust}}$ values that are too low for the case of 1 and 2\,$M_\odot$ at 5\,Myr (both dust mass: directly from models or the synthetic dust mass as inferred from the simulations). 

In the case of weak traps, we obtained two interesting results. First, when comparing it with the mass in large dust directly obtained from the models, the $M_{\rm{dust}}-M_{\star}$ relation of TDs and disks with sub-structures is reproduced at 1\,Myr of evolution. The sample of TDs and disks with substructures (Table~\ref{table:disk_structures} in the appendix) is a combination of disks at different stages of evolution, from 0.5 to 10\,Myr, but most of them are older than 1\,Myr \citep{garufi2018}. Second, due to the reasons explained above (optical thickness and boulder formation), when comparing it with the synthetic dust mass, we find that the model becomes steeper and closer to the slope of different star forming regions, especially at 1\,Myr. At 5\,Myr, the synthetic dust mass is lower than in any star forming region, in particular for disks around 1 and 2\,$M_\odot$ stellar masses. 

The case of strong traps shows very interesting results but they must be interpreted with caution. First, the $M_{\rm{dust}}-M_{\star}$ relation observed in TDs and disks with substructures is fairly reproduced at any time of evolution when comparing with the mass in large dust obtained directly from the models. However, due to boulder formation, in particular for the 0.1 and 0.3\,$M_\odot$, the synthetic dust mass is much lower reproducing again the steeper trends at any time of evolution. This implies that to explain the trend of TDs and disks with substructures, strong pressure bumps (perhaps due to giant planets) are needed, but in addition, boulder formation must be inhibited inside these pressure bumps, otherwise the observed fluxes and masses are too low, reproducing instead the steeper relations in different star-forming regions.

Therefore, in the context of planets causing these pressure bumps, we found similar results as in the case of $M_{\rm{disk}}=0.05\,M_\odot$ for the disks around 1 and 2\,$M_\odot$, meaning those for which only strong pressure bumps (or giant planets) can explain the observed values of dust mass at any time of evolution. Contrary to the case of $M_{\rm{disk}}=0.05\,M_\odot$, any of the three scenarios (no traps, weak or strong traps) can reproduce the $M_{\rm{dust}}-M_{\star}$ relations of different star formation regions for 0.1 and 0.3\,$M_\odot$ at 1 or 5 Myr of evolution. The agreement with the strong traps scenario is surprising, but it is due to boulder formation, while in the other two cases, it is due to the actual loss of particles attributed to the drift. 

\begin{figure*}
 \centering
        \centering
        \includegraphics[width=18cm]{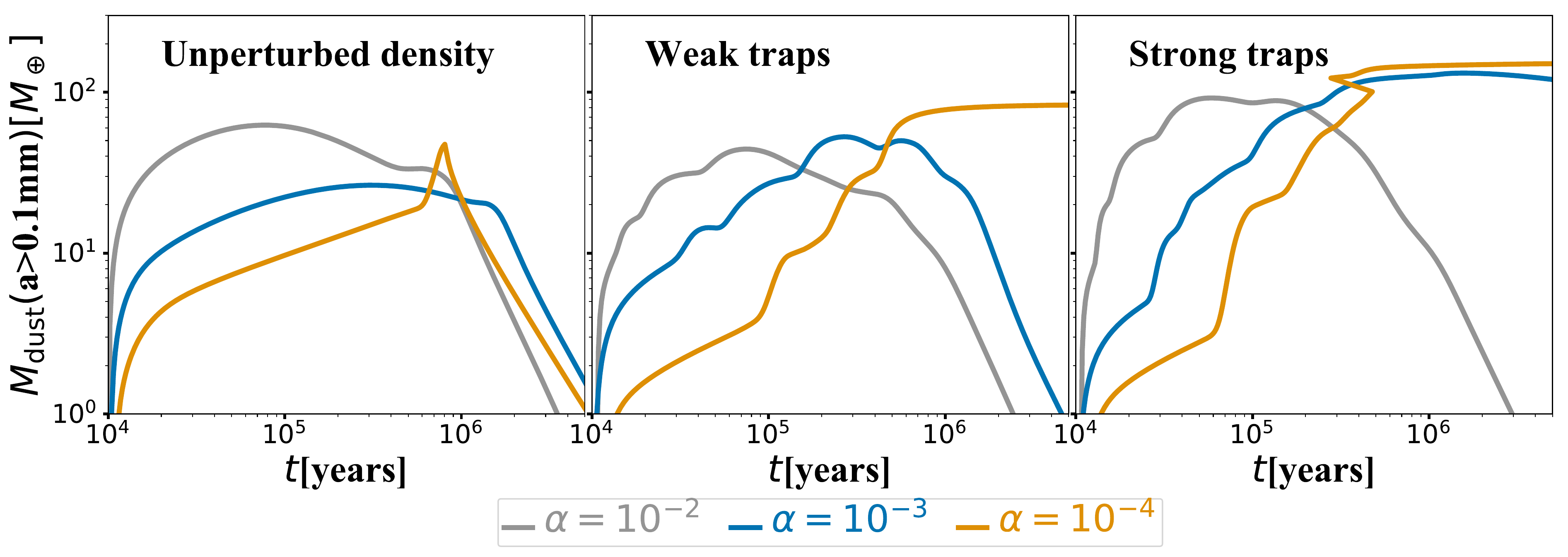}
   \caption{Effect of the turbulence mixing strength on the evolution of the dust mass of particles larger than 0.1mm from the models with no traps (left panel), weak traps (middle panel), and strong traps (right panel). In all the cases, $M_\star=0.1\,M_\odot$ and $M_{\rm{disk}}=0.05\,M_\odot$.}
   \label{alpha_effect}
\end{figure*}

\section{Discussion} \label{sect:discussion}

\subsection{Balance between trapping and boulder formation}
Our findings suggest that strong pressure traps can reproduce the $M_{\rm{dust}}-M_{\star}$ relation observed in different star forming regions (bottom right panel of Fig.~\ref{Mstar_Mdisks_Mdisk}) because of the trapping of dust particles and boulder formation inside pressure bumps, the latter in particular for $M_\star<1\,M_\odot$. In contrast, the flatter relation observed for TDs and disks with substructures has not been reproduced by any model,  specifically one based on post-processing for the purpose of comparison with the observations. In the framework of our models, one solution to reproduce the relation of TDs and disks with substructures is to consider that boulder formation is inhibited inside pressure bumps in the case of low-mass stars ($M_\star<1\,M_\odot$). 

Our results seem to contradict those of \cite{stammler2019}, who suggests that dust must be transformed into planetesimals to obtain the optical depths values inferred inside the DSHARP rings \citep{andrews2018}, using dust evolution models that include trapping in a pressure bump. However, that work was done assuming a Herbig star, in particular to reproduce the observations of one of the brightest disk (HD\,163296), while our conclusion about hindering boulder formation inside pressure bumps applies only to low-mass stars ($M_\star<1\,M_\odot$). In our case of a disk around a Herbig star, we do not have any natural production of boulders in pressure maxima and the small differences (a factor of $\sim1.5-2$) between the mass in large dust directly from the models and the observed total dust mass originate from the high optical depth ($\sim$1-2) inside the pressure maxima. According to our results, inducing boulder formation in the pressure bumps in the Herbig case can help to  reduce $\tau$ and be in agreement with the values obtained in the DSHARP sample. However, as found in our models, boulder formation reduces by a lager factor the synthetic dust mass ($\sim10-100$, comparison between right panels of Fig.~\ref{Mstar_Mdisks_Mdisk}). Assuming a higher initial disk mass may help to have a better agreement with the synthetic dust mass as in \cite{stammler2019}, where the initial condition for the disk mass is 0.4\,$M_\odot$ (four times than what we assume in this work) and they do reproduce the intensity of one of the observed rings of HD\,163296.

There are several possibilities that may hinder boulder formation in pressure maxima in disks around low-mass stars. The first possibility is that the velocity threshold for which particles fragment (fragmentation velocity) is lower than what we considered ($10$\,ms$^{-1}$). Laboratory experiments and numerical simulations show that these values are possible \citep[e.g.,][]{paszun2009, gundlach2015, musiolik2016} for water ice grains, although recent experiments suggest that the fragmentation velocities for water ice grains may be as low or lower as for silicates, that is, $1$\,ms$^{-1}$ \citep[e.g.,][]{musiolik2019}. However, if this is the case, the fragmentation barrier would be two orders of magnitude lower (Eq.~\ref{afrag}), which will imply that the maximum grain size would be set by fragmentation with typical values of $\sim$100$\mu$m in the outer disk ($>$20\,au). These grain sizes are in agreement with the values proposed by \cite{liu2019} and  \cite{zhu2019}, who suggest that the measured optical depth  in several DSHARP disks can be explained by dust scattering when the optically thick dust has an albedo of $\sim$0.9 at 1.25\,mm, which corresponds to a maximum grain size of $\sim$0.1-1\,mm. Nevertheless, if this is the case, dust trapping may be inefficient in reproducing observations. Investigating the required conditions  for  particle trapping, and an agreement with observations when the fragmentation velocities are $1$\,ms$^{-1}$ in the entire disk, is a subject of future research.

Other possibilities that may explain the inhibition of the formation of boulders  include bouncing in the models or an increase in the disk $\alpha$ viscosity. Nevertheless, to include bouncing would keep the grains too small \citep[Stokes numbers lower than $10^{-4}$, e.g.,][]{zsom2010} such that they could not effectively be trapped in pressure maxima. The other alternative is to have a higher value of the $\alpha$ viscosity, which sets the fragmentation barrier and turbulence mixing strength; hence, determining how much grains can be diffused out from pressure bumps, which is discussed in the next subsection.

\subsection{Effect of turbulence mixing strength}
As discussed in the previous subsection, increasing the turbulence mixing strength may help to hinder boulder formation inside pressure bumps. In general, changing the value of the $\alpha$ viscosity in our models has a strong effect in the results. This effect has been studied and discussed in detail in \cite{ovelar2016} and \cite{bae2018}, who show that by just varying the values of $\alpha$ and keeping the rest of the model parameters the same, the synthetic images from models can have very different outputs, such as: multiple rings and gaps, a cavity with a single ring, or compact disks. 

\begin{figure*}[!t]
   \tabcolsep=0.05cm 
   \begin{tabular}{cc}
   \centering
    \includegraphics[width=\columnwidth]{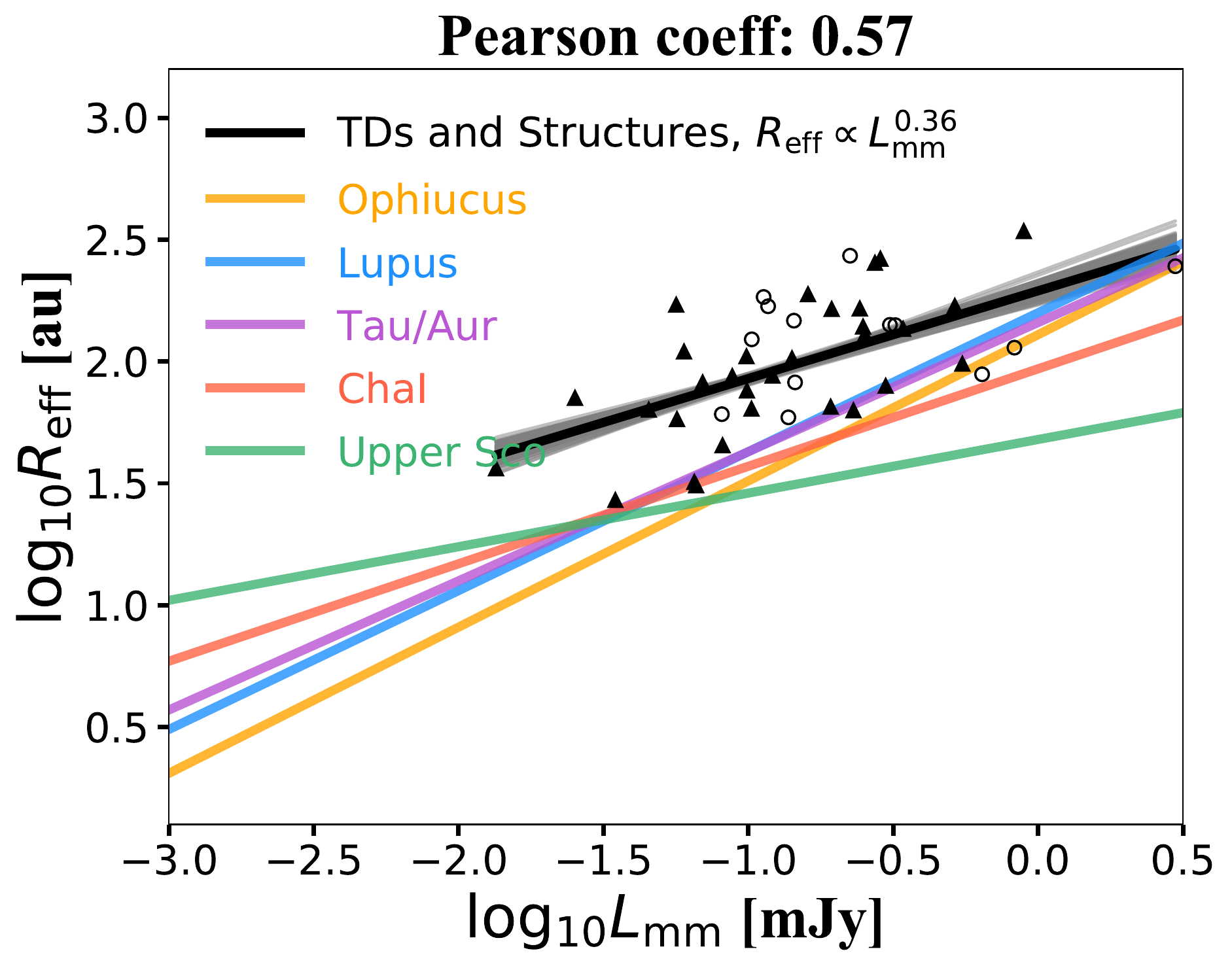}&
    \includegraphics[width=\columnwidth]{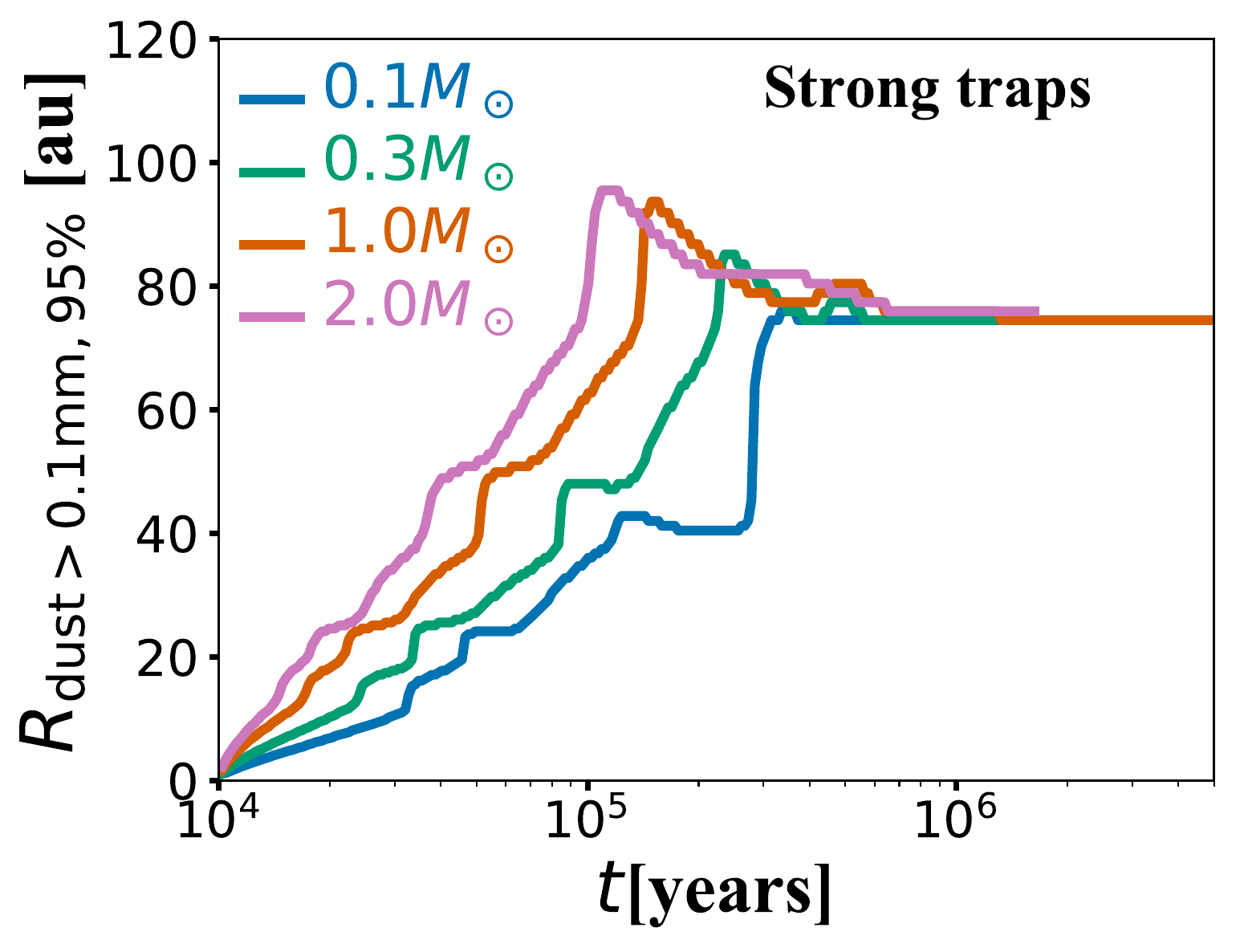}
    \end{tabular}
    \caption{Left panel: fit of the disk size-luminosity relation from sub-millimeter observations of protoplanetary disks in different star forming regions (colors) as reported in \cite{hendler2020} compare to the relation for TDs (open circles) and disks with substructures (full triangles). Right panel: evolution of the radius that encloses 95\% of the mass of dust particles larger than 0.1\,mm from the models that assume strong pressure bumps, $M_{\rm{disk}}=0.05\,M_\star$, and $\alpha=10^{-3}.$ }
    \label{fig:disk_size}
\end{figure*}

In the case of planets, usually a higher mass planet is required to open a gap when the disk viscosity is higher. This balance between planet mass and viscosity has an effect in the potential trapping of particles \citep[Fig.~1 in][]{ovelar2016}. The amplitude of $A=4$  of our perturbation in Eq.~\ref{Gaussian_perturbation} corresponds to planet masses similar to 0.1-1\,$M_{\rm{Jup}}$ for $10^{-3}-10^{-2}$ $\alpha-$viscosity. We note that increasing the amplitude of our perturbation, which would resemble gaps formed by more massive planets or less viscous disks, does not have a strong effect in our models, since with $A=4$ effective trapping is already reached and therefore increasing $A$ would only saturate the amount of dust trapped in pressure maxima. In our assumptions of the particle traps, the parametric perturbation does not depend on $\alpha$, and hence the shape of the pressure bumps is independent of gas viscosity. As a consequence, the only effect of varying $\alpha$ in our models is reflected in the dust evolution. 
 
Varying $\alpha$ directly influences the fragmentation barrier $a_{\rm{frag}}\propto1/\alpha$, which means that for higher $\alpha$, grains reach lower sizes when the maximum grain size is set by fragmentation. In addition, we assume dust diffusion $D$ as \cite{youdin2007}, which means $D\propto\alpha/(1+St^2)$ and, thus, higher dust diffusion for higher values of $\alpha$. Increasing $\alpha$ implies lower $a_{\rm{frag}}$ and higher $D$, which influence the potential trapping because if grains do not grow to values that attain a high Stokes number, they are more difficult to trap and they diffuse out of pressure traps easily. To see the effect of $\alpha$ on the mass in large dust, we simulated the case of  $M_\star=0.1\,M_\odot$ and $M_{\rm{disk}}=0.05\,M_\odot$ with three different values of $\alpha$ ($10^{-4}$, $10^{-3}$, $10^{-2}$). Figure~\ref{alpha_effect} shows the main results of this numerical experiment. 
 
When the disk viscosity is high ($\alpha=10^{-2}$) trapping is less efficient and it does not occur even for the case of strong traps (Fig.~\ref{alpha_effect}). In these cases, the peak of  the mass in large dust is reached earlier in the simulations ($\sim10^{5}$ years). On the other hand, when  disk viscosity is  low ($\alpha=10^{-4}$),  the peak of  the mass in large dust is reached later ($\sim$1\,Myr), and in this case trapping is very efficient even for weak traps. In this case, the fragmentation barrier is one order of magnitude higher than our standard models, which implies more efficient formation of boulders inside pressure bumps, creating more invisible solid material when observed at millimeter-emission and a higher discrepancy with the observed dust mass. 

\subsection{Disk size-luminosity relation}
\cite{tripathi2017} and \cite{andrews2018a} find a significant correlation between the disk size ($R_{\rm{eff}}$, usually defined as the radii that encloses 68\% of the millimeter emission) and the continuum luminosity of the disk. \cite{hendler2020} performed a homogeneous analysis of five star forming regions, finding that this relation flattens for Upper Sco, which is the oldest region in the sample (left panel of Fig.~\ref{fig:disk_size}). The observed relation could originate from the growth and radial drift of dust particles, where the dust disk extension becomes smaller with time due to drift. Depending on the initial conditions, dust evolution models can reproduce the observed relations \citep{tripathi2017, rosotti2019}. It is, therefore, possible that the $R_{\rm{eff}}-L_{\rm{mm}}$ flattens for disks in which radial drift is reduced or ineffective. 

We checked the $R_{\rm{eff}}-L_{\rm{mm}}$ relation for our sample of TDs and disks with substructures and found a significan correlation (Pearson coefficient of 0.57) with an intercept (A) equal to $2.29\pm0.02$ and a slope (B) equal to $0.36\pm0.03$, assuming $\log_{10}(R_{\rm{eff}}/\rm{au})=B \log_{10}(L_{\rm{mm}}/\rm{mJy})+A$. For this calculation, the millimeter luminosity is scaled to a 140\,pc distance for all the disks. The values for $R_{\rm{eff}}$ are the ones that enclose 90 or 95\% of the millimeter flux in Band 6 or Band 7 with ALMA (the wavelength is between $\sim0.8$ and $\sim$1.3\,mm).  The relations found in \cite{hendler2020}, \cite{tripathi2017} and \cite{andrews2018a} used the radius that encloses the 68\% of the flux, but \cite{hendler2020} demonstrates that there is a linear relation close to 1-1 between the two values.

The values for $R_{\rm{eff}}$ of the sample are larger than the averaged values found in different star forming regions, with a mean value of $116$\,au (see Table\,2 from \cite{hendler2020} for comparison). This is because most of the ALMA observations of TDs and disks with substructures are  biased towards disks with larger radii. The slope of $R_{\rm{eff}}-L_{\rm{mm}}$ in our sample is smaller than the values found in the youngest star forming regions studied in \cite{hendler2020}, that is, Ophiucus, Lupus, and  Taurus/Auriga with ages between 1 and 3\,Myr. The slope of our sample lies in between the values found for Chamaeleon I (2-3 Myr, slope of 0.4) and Upper-Sco (5-1\,Myr, slope of 0.22), while the ages of the TDs and disks with substructures in our sample range between $0.1\,$\,Myr and 10\,Myr.  

Based on our dust evolution models, we calculate the radius that encloses 95\% of the mass of the dust particles that are larger than 0.1\,mm. The evolution of this radius is shown in the right panel of Fig.~\ref{fig:disk_size} for the models that assume strong pressure bumps, $M_{\rm{disk}}=0.05\,M_\star$, and $\alpha=10^{-3}$. Independent of the stellar mass, after $\sim$0.5\,Myr of evolution the value of this radius converges to the location of the farther pressure maximum, that is, $\sim75\,$au (similar values are found when the disk radius is calculated from the synthetic intensity profile). Based on these models, we expect a flat relation between  $R_{\rm{eff}}$ and $L_{\rm{mm}}$ at any evolutionary stage. It is possible that the observed $R_{\rm{eff}}-L_{\rm{mm}}$ relation for TDs and disks with substructures provides information about how far pressure bumps may form and how the less luminous disks that are around stars of lower mass may have the pressure bumps closer in. Exoplanet statistics suggest that low-mass stars have more planets with radii of 1-4\,$R_\oplus$ at the same semi-major axis \citep[e.g.,][]{mulders2015} but to test this idea, we require more detailed models to investigate different locations and the morphology of pressure bumps, along with a greater number of high-resolution observations around low-mass stars.

\section{Conclusions} \label{sect:conclusion}

We investigated the effect of dust evolution and particle trapping in the observed  $M_{\rm{dust}}-M_{\star}$ relation of different star-forming regions. We focused our attention on the effect of the stellar and disk mass. Both properties affect the mass in large dust at million-year timescales due to: (1) the more effective drift around low-mass stars; (2) the more efficient fragmentation for disks around stars of higher mass - which are less efficient in more massive disks; and (3) the drift being more efficient in disks of lower mass.  The combination of these effects influence the efficiency of trapping and the potential to form boulders in pressure bumps, which influence the final  $M_{\rm{dust}}$. Our conclusions are:

\begin{enumerate}

    \item Independent of the assumption for the disk mass, either  $M_{\rm{disk}}=0.05\,M_\odot$ or $M_{\rm{disk}}=0.05\,M_\star$, strong traps are required to reproduce the observed  $M_{\rm{dust}}$ of disks around  $M_\star\geq 1\,M_\odot$ at different time of evolution. In the context of planets as the origin of pressure bumps, this result agrees with current exoplanet statistics about giant planets, which are found to be more common around more massive stars.

    \item When all disks have the same initial mass independent of the stellar mass ($M_{\rm{disk}}=0.05\,M_\odot$), the dust evolution models produce a flatter $M_{\rm{dust}}-M_{\star}$ relation when compared to the observed ones from different star-forming regions. Therefore, we  conclude that the initial disk mass has to scale with the stellar mass to reproduce the slope of the $M_{\rm{dust}}-M_{\star}$  relation, otherwise it is flatter than what is observed. In the case of $M_{\rm{disk}}=0.05\,M_\odot$ for all cases, the slope obtained from the models is similar to the one observed for TDs and disks with substructures, however the modeled $M_{\rm{dust}}$ is overall lower than the observed one.

    \item When the disk mass is a fraction of the stellar mass (5\%), our models reproduce the observed $M_{\rm{dust}}-M_{\star}$ relation from different star-forming regions. However, in the cases of an unperturbed density and weak traps, the trends are reproduced only at 1\,Myr while at later times (5\,Myr), $M_{\rm{dust}}$ is too low when compared to observations for $M_\star\geq1M_\odot$.

    \item Strong pressure traps can reproduced the observed $M_{\rm{dust}}-M_{\star}$ relations of different star forming regions in the case of $M_{\rm{disk}}=0.05\,M_\star$. This result arises from dust trapping and dust growth beyond centimeter-sizes inside pressure traps, an effect that decreases the millimeter fluxes and reduces the expected $M_{\rm{dust}}$, in particular for $M_\star<1\,M_\odot$.

    \item To explain the flatter relation of $M_{\rm{dust}}-M_{\star}$ for TDs and disks with substructures, strong pressure bumps (perhaps caused by giant planets) are needed. However, efficient growth must be inhibited inside these pressure bumps for low-mass stars, otherwise the observed fluxes and masses are too low, reproducing instead the steeper trends in different star-forming regions. Different possibilities can hinder very efficient growth inside pressure bumps, such as a lower fragmentation velocity or including the effect of bouncing.
    
    \item Due to the efficient formation of decimenter bodies in pressure bumps for disks around low-mass stars $M_\star<1\,M_\odot$, we cannot give a definitive conclusion on what type of traps are present in these disks. Indeed,  all three cases we tested (unperturbed, weak, and strong traps) can reproduced the observed values in the more realistic case of disk mass scaling with stellar mass. 
    
    \item The $R_{\rm{eff}}-L_{\rm{mm}}$ relation for TDs and disks with substructures is flatter than observed in Ophiucus, Lupus, and  Taurus/Auriga star-forming regions and the slope value of our sample lies in the range found for older regions, that is, Chamaeleon I and Upper-Sco. This relation may flatten due to inefficient radial drift when pressure bumps exist in the disks, in which case $R_{\rm{eff}}$ traces the location of the farther pressure bump at any time of evolution. Currently, high angular resolution observations of TDs and disks with sub-structures are biased toward the brightest and larger disks, so more observations of small and faint ones are required to test this idea. 
    
\end{enumerate}

Future multi-wavelength and high-resolution observations of disks around low-mass stars are needed to investigate if gaps or rings are present and how much grains have growth there in comparison to the rings and gaps in disks around more massive stars. Our results suggest that boulder formation must be inhibited inside pressure bumps in TDs and disks with sub-structures in order to explain the observed $M_{\rm{dust}}$. Further theoretical research is needed to investigate what conditions are required to explain the observed properties of this group of disks, in particular if boulder formation is hindered when the fragmentation velocities are lower than assumed in this work - as has been suggested by recent laboratory experiments.

\section*{Acknowledgements}
The authors are thankful to the referee for the constructive referee report, to F. Long for all the enthusiastic discussions about the results of this paper, and to N. van der Marel for sharing some data. P.P. acknowledges support provided by the Alexander von Humboldt Foundation in the framework of the Sofja Kovalevskaja Award endowed by the Federal Ministry of Education and Research. I.P. acknowledges support from an NSF Astronomy \& Astrophysics Research Grant (ID: 1515392). This material is based on work supported by the National Aeronautics and Space Administration under Agreement No. NNX15AD94G for the program Earths in Other Solar Systems. The results reported herein benefited from collaborations and/ or information exchange within NASA’s Nexus for Exoplanet System Science (NExSS) research coordination network sponsored by NASA’s Science Mission Directorate.

\bibliographystyle{aa} 
\bibliography{pinilla_Must_Mstar.bbl}

\begin{appendix}
\section{Target list of TDs and disks with substructures}

\begin{table*}
\centering   
\begin{tabular}{|c|c|c|c|c|c|c|c|}
\hline
\hline       
\textbf{Target}&
d&
$\log_{10}M_\star$&
$\lambda$&
$F_{\rm{mm}}$&
$R_{\rm{90-95\%}}$&
TD&
Ref\\
& 
[pc]&
[$M_\odot$]&
[mm]&
[mJy]&
[au]&
&
\\
\hline
AS\,209  &121& -0.08 &1.25&288.0&139.0&\xmark&1,2\\
DoAr\,25 &138& -0.02 &1.25&246.0&165.0&\xmark&1,2\\             
DoAr\,33 &139& 0.04  &1.25&35.0&27.0&\xmark&1,2\\            
Elias\,20&138& -0.32 &1.25&104.0&64.0&\xmark&1,2\\                
Elias\,24&136& -0.11 &1.25&352.0&136.0&\xmark&1,2\\           
Elias\,27&116& -0.31 &1.25&330.0&254.0&\xmark&1,2\\           
GW\,Lup  &155& -0.34 &1.25&89.0&105.0&\xmark&1,2\\           
HD\,142666&148&0.20  &1.25&130.0&59.0&\cmark&1,2\\          
HD\,144006&165&0.25  &1.25&59.0&82.0&\xmark&1,2\\           
HD\,163296&101&0.31  &1.25&715.0&169.0&\xmark&1,2\\          
IM\,Lup  &158&-0.05  &1.25&253.0&264.0&\xmark&1,2\\            
MY\,Lup  &156&0.09   &1.25&79.0&87.0&\xmark&1,2\\             
RU\,Lup  &159&-0.20  &1.25&203.0&63.0&\xmark&1,2\\            
SR\,4    &134&-0.17  &1.25&69.0&31.0&\xmark&1,2\\             
Sz\,114  &162&-0.76  &1.25&49.0&58.0&\xmark&1,2\\           
Sz\,129  &161&-0.08  &1.25&86.0&76.0&\xmark&1,2\\            
WaOph    &123&-0.17  &1.25&161.0&103.0&\xmark&1,2\\           
WSB\,52  &136&-0.32  &1.25&67.0&32.0&\xmark&1,2\\
HT\,Lup  &154&0.23   &1.25&77.0&--&\xmark&1,2\\
AS\,205  &127&-0.06  &1.25&358.0&--&\xmark&1,2\\
CI\,Tau  &158&-0.05  &1.30&142.4&188.8&\xmark&3\\             
CIDA\,9A &171&0.12   &1.30&37.1&63.4&\cmark&3\\             
DL\,Tau  &159&0.01   &1.30&170.7&164.2&\xmark&3\\             
DN\,Tau  &128&-0.06  &1.30&88.6&60.8&\xmark&3\\             
DS\,Tau  &159&-0.08  &1.30&22.2&70.9&\xmark&3\\             
FT\,Tau  &127&-0.47  &1.30&89.8&45.3&\xmark&3\\             
GO\,Tau  &144&-0.31  &1.30&54.8&170.9&\xmark&3\\             
IP\,Tau  &130&-0.03  &1.30&14.5&36.4&\cmark&3\\             
IQ\,Tau  &131&-0.13  &1.30&64.1&109.8&\xmark&3\\                    
MWC\,480 &161&0.32   &1.30&267.8&141.4&\xmark&3\\             
RY\,Tau  &128&0.31   &1.30&210.4&65.2&\cmark&3\\             
UZ\,Tau E&131&0.09   &1.30&129.5&87.4&\cmark&3\\           
AA\,Tau  &137&0.02   &1.10&105.0&123.3&\cmark&4\\                          
DM\,Tau  &145&0.13   &1.10&109.0&184.0&\cmark&4\\                          
HL\,Tau  &147&-0.13  &1.30&789.0&114.0&\xmark&4\\                       
Elias\,24&136&-0.07  &1.30&331.0&141.0&\xmark&4\\                       
GY\,91   &137&-0.46  &0.86&258.0&128.8&\xmark&4\\                          
HD\,100546&110&0.38  &1.30&379.0&79.4&\cmark&4\\                          
HD\,135344B&136&0.16 &0.85&564.0&97.9&\cmark&4\\                        
HD\,169142&114&0.24  &1.30&178.0&82.1&\cmark&5\\                    
HD\,97048&185&0.34   &0.90&2253.0&246.0&\cmark&6\\                   
RXJ\,1615&158&0.06   &0.44&878.0&139.0&\cmark&4\\                         
Sz\,98   &156&0.06   &1.30&105.0&168.5&\xmark&4\\            
TW\,Hya  &60&-0.12   &0.87&1495.0&88.6&\cmark&4\\              
V1094\,Sco&154&0.02  &1.30&204.0&271.7&\xmark&4\\          
V1247\,Ori&398&0.29  &0.85&314.0&342.7&\cmark&4\\          
J16083070&156&0.20   &0.89&128.9&147.1&\cmark&7\\          
RY\,Lup &159&0.14    &0.89&263.9&118.6&\cmark&7\\            
Sz\,111 &158&-0.32   &0.89&176.7&123.6&\cmark&7\\            
Sz\,100 &137&-0.83   &0.89&53.6&65.6&\cmark&7\\             
J160708 &176&-0.76   &0.89&85.0&184.7&\cmark&7\\             
\hline
\hline
\end{tabular} 
\vspace{0.3cm}
\caption{Targets, distance, stellar mass, observed wavelength, fluxes, and a radii that encloses 90 or 95\% of the millimeter flux from ALMA observations of TDs and disk with substructures. The TD column is \cmark  for disk classified as TDs. All distances from Gaia \citep[except for Sz123A, J10581677-7717170,][]{gaia2016, gaia2018} References: (1)\cite{andrews2018}, (2)\cite{huang2018}, (3)\cite{long2018}, (4)\cite{marel2019}, (5)\cite{sebaperez2019}, (6)\cite{pinte2019}, (7)\cite{pinilla2018}, (8)\cite{cieza2019}, (9)\cite{macias2018}, (10) \cite{kastner2018}, and (11)\cite{dong2018}.}
\label{table:disk_structures}
\end{table*}

\begin{table*}
\centering   
\begin{tabular}{|c|c|c|c|c|c|c|c|}
\hline
\hline       
\textbf{Target}&
d&
$\log_{10}M_\star$&
$\lambda$&
$F_{\rm{mm}}$&
$R_{\rm{90-95\%}}$&
TD&
Ref\\
& 
[pc]&
[$M_\odot$]&
[mm]&
[mJy]&
[au]&
&
\\
\\
\hline
Sz\,118 &164&-0.04   &0.89&59.7&96.1&\cmark&7\\      
Sz\,123A&200&-0.29   &0.89&39.7&60.6&\cmark&7\\      
J16042165&150&0.08   &1.30&69.1&120.6&\cmark&7\\           
J10581677&160&0.10   &0.89&330.0&207.7&\cmark&7\\          
J10563044&183&-0.07  &0.89&141.9&158.2&\cmark&7\\          
DoAr\,44 &164&0.11   &0.89&180.4&61.1&\cmark&7\\     
LkCa\,15 &159&0.00   &0.44&1458.1&138.1&\cmark&7\\
SR\,21   &138&0.29   &0.87&347.0&61.2&\cmark&7\\ 
SR\,24S  &114&-0.09  &1.30&227.2&79.6&\cmark&7\\             
Sz\,91   &159&-0.31  &0.89&34.3&173.2&\cmark&7\\             
TCha     &107&0.03   &0.89&225.2&63.1&\cmark&7\\               
HD\,34282&312&0.27   &0.85&333.7&253.3&\cmark&7\\               
CIDA\,1  &136&-0.96  &0.89&35.4&27.0&\cmark&7\\             
CQ\,Tau  &163&0.17   &1.30&172.2&69.1&\cmark&7\\             
UXTau\,A &140&0.14   &1.30&65.0&46.7&\cmark&7\\             
V892\,Tau&117&0.44  &1.30&286.7&50.1&\cmark&7\\             
$\rho$Oph\,3&161&--  &1.30&96.4&38.5&\cmark&7\\
$\rho$Oph\,38&156&-- &1.30&191.5&--&\cmark&8\\
RXJ\,1633.9-2442&135&--&1.30&79.8&--&\cmark&8\\
ROXRA\,3&143.7&--    &1.30&72.3&--&\cmark&8\\
IRAS\,16201-2410&160.5&--&1.30&43.4&--&\cmark&8\\
GM\,Aur&160&0.04&0.90&380&--&\cmark&9\\
V4046\,Sgr&72&-0.05&0.95&472.0&60.0&\cmark&10\\
MWC\,758&160&0.15&0.86&1445.0&--&\cmark&11\\
\hline
\hline
\end{tabular}   
\vspace{0.3cm}
\caption{Continuation Table~\ref{table:disk_structures}.}
\end{table*}
\end{appendix}
\end{document}